\documentclass{aa_arxiv}
\usepackage[colorlinks,citecolor=blue,linkcolor=blue,urlcolor=blue]{hyperref}
\usepackage{array}
\usepackage[normalem]{ulem}
\newcommand{\graphreduce}{0.95}

\title{Chaotic magnetic disconnections trigger flux eruptions in~accretion flows channeled onto magnetically saturated Kerr~black~holes}
\titlerunning{Chaotic disconnections trigger black hole flux eruptions}

\author{
Krzysztof~Nalewajko\inst{\ref{inst_camk}}\thanks{\email{knalew@camk.edu.pl}}
\and
Mateusz~Kapusta\inst{\ref{inst_oauw}}
\and
Agnieszka~Janiuk\inst{\ref{inst_cft}}
}
\authorrunning{Nalewajko, Kapusta \& Janiuk}

\institute{
Nicolaus Copernicus Astronomical Center, Polish Academy of~Sciences, Bartycka 18, 00-716 Warsaw, Poland
\label{inst_camk}
\and
Astronomical Observatory, University of Warsaw, Al. Ujazdowskie 4, 00-478 Warsaw, Poland
\label{inst_oauw}
\and
Center for Theoretical Physics, Polish Academy of Sciences, Al. Lotnikow 32/46, 02-668 Warsaw, Poland
\label{inst_cft}
}

\abstract
{
Magnetized accretion flow onto a black hole (BH) may lead to accumulation of poloidal magnetic flux across its horizon, which for high BH spin can power far-reaching relativistic jets.
The BH magnetic flux is subject to a saturation mechanism by means of magnetic flux eruptions involving relativistic magnetic reconnection.
Such accretion flows have been described as magnetically arrested disks (MAD) or magnetically choked accretion flows (MCAF).
}
{
The main goal of this work is to describe the onset of relativistic reconnection and initial development of magnetic flux eruption in accretion flow onto magnetically saturated BH.
}
{
We analyze the results of 3D general relativistic ideal magnetohydrodynamic (GRMHD) numerical simulations in the Kerr metric, starting from weakly magnetized geometrically thick tori rotating either prograde or retrograde.
We integrate large samples of magnetic field lines in order to probe magnetic connectivity with the BH horizon.
}
{
The boundary between magnetically connected and disconnected domains coincides roughly with enthalpy equipartition.
The geometrically constricted innermost part of the disconnected domain develops a rigid structure of magnetic field lines -- rotating slowly and insensitive to the BH spin orientation.
The typical shape of innermost disconnected lines is a double spiral converging to a sharp inner tip anchored at the single equatorial current layer.
The footpoints of magnetic flux eruptions are found to zip around the BH along with other azimuthal patterns.
}
{
Magnetic flux eruptions from magnetically saturated accreting BHs can be triggered by minor density gaps in the disconnected domain, resulting from chaotic disconnection of plasma-depleted magnetospheric lines.
Accretion flow is effectively channeled along the disconnected lines towards the current layer, and further towards the BH by turbulent cross-field diffusion.
Rotation of flux eruption footpoints may contribute to the variability of BH crescent images.
}


\begin{document}
\maketitle

\section{Introduction}

Accretion onto black holes (BH) is the engine of various astrophysical phenomena, ranging from active galactic nuclei (AGN), tidal disruption events (TDE), gamma-ray bursts (GRB), X-ray binaries (XRB).
Accretion flows have long been considered to be magnetized \citep{1969Natur.223..690L,1973A&A....24..337S}, and even in very weakly magnetized accretion flows the magnetic field can play fundamental roles, e.g., magnetorotational instability (MRI) enables accretion by transporting angular momentum outwards \citep{1991ApJ...376..214B}.
Accretion flows can also transport magnetic fields towards the BH, depending on the effective balance between advection and diffusion \citep{1994MNRAS.267..235L}.
Advection of magnetic fields is more efficient for geometrically thick accretion flows; at low mass accretion rates (in terms of the Eddington luminosity) such flows are hot, radiatively inefficient (RIAF), and (heat-)advection dominated (ADAF) \citep{2014ARA&A..52..529Y}.

As plasma is swallowed by the BH, it is separated from poloidal magnetic field, which accumulates across the BH outer horizon with unsigned magnetic flux $\Phi_{\rm BH}$, easily overcoming the gravitational expulsion (Meissner) effect (e.g., \citealt{2007MNRAS.377L..49K}; \citealt{2024A&A...687A.185J}).
The magnetic fields connected to the BH horizon form a highly magnetized, essentially force-free, bipolar BH magnetosphere.
If the BH is spinning, such a magnetosphere induces an outward Poynting flux in the Blandford-Znajek mechanism \citep{1977MNRAS.179..433B}, electromagnetic equivalent of the Penrose process \citep{Lasota2014}.
Such Poynting flux is proportional to $\Phi_{\rm BH}^2$ and a function of the BH spin $a$ \citep{Tchekhovskoy2010,2022JCAP...07..032C}.
It drives a powerful outflow that gradually accelerates and collimates into a pair of relativistic jets (\citealt{1984RvMP...56..255B}; \citealt{2019ARA&A..57..467B}).

Relativistic jets have dramatic observational manifestations.
They are directly observed in $\sim 10\%$ of AGN belonging to the radio-loud group \citep[e.g.,][]{1998AJ....115.1295K}, which includes blazars (with huge relativistic luminosity boost due to jet orientation close to the line of sight) and radio galaxies (without strong luminosity boost) \citep{1995PASP..107..803U}; and also in certain low-mass XRB (microquasars, \citealt{1999ARA&A..37..409M}).
Moreover, they are inferred indirectly due to inevitability of relativistic beaming in all GRB \citep{2004RvMP...76.1143P,2018pgrb.book.....Z} and some TDE \citep{2021ARA&A..59...21G}.
Many relativistic jets dissipate very significant fractions of their power \citep{2014Natur.515..376G} and are very efficient particle accelerators (\citealt{2020NatAs...4..124B}; \citealt{2023Sci...380.1390L}), which makes blazars the dominant cosmic sources of energetic gamma-ray radiation (\citealt{2015ApJ...800L..27A}; \citealt{2016ARA&A..54..725M}).
The most detailed observational picture of a relativistic jet, in particular of its gradual collimation and origin at the BH, is provided by one of the nearest radio galaxies, M87 (e.g., \citealt{1999Natur.401..891J,2016ApJ...817..131H,2018ApJ...855..128W,2018A&A...616A.188K}; \citealt{2023Natur.616..686L}).

A higher $\Phi_{\rm BH}$ allows for a more powerful jet, but it also means a stronger backreaction of the BH magnetosphere on the accretion flow.
There is an upper limit on the value of $\Phi_{\rm BH}$ (and hence on the power of relativistic jets driven by Poynting fluxes), roughly $\Phi_{\rm BH} \lesssim 50 \dot{M}_{\rm acc}^{1/2}$ (in the natural Gaussian units with $c = G = 1$) with $\dot{M}_{\rm acc}$ the mass accretion rate \citep{2011MNRAS.418L..79T}, with a moderate dependence on the BH spin \citep{2022MNRAS.511.3795N,2024ApJ...962..135Z}.
Accretion flows interacting with such saturated magnetospheres have been identified by \cite{2011MNRAS.418L..79T} as the Magnetically Arrested Disks (MAD), referring to a model proposed by \cite{2003PASJ...55L..69N}\footnote{Accretion flow would be arrested by magnetic field perpendicular to the flow velocity and forming a direct barrier with sharp radial stratification at a particular magnetospheric radius $R_{\rm m}$.}.
Alternatively, they have been described by \cite{2012MNRAS.423.3083M} as the Magnetically Choked Accretion Flows (MCAF)\footnote{\cite{2012MNRAS.423.3083M} define magnetic choking as compression of the accretion flow by magnetic pressure of the BH magnetosphere, `analogous to chokes in man-made engines'.}.
In Section \ref{sec_disc}, we argue that at magnetic saturation the bulk of accretion flow is effectively `channeled' along regular magnetic fields and the equatorial current layer.

The saturation mechanism for BH magnetic flux involves episodic, relatively fast decreases of $\Phi_{\rm BH}$ with amplitudes up to roughly 50\% of the preceding maximum levels.
Associated with such BH flux decreases are magnetic flux eruptions localized azimuthally (along coordinate $\phi$), directed radially outwards along the equatorial plane, resulting in compact tubes of largely `vertical' (perpendicular to the equatorial plane) magnetic fields ejected from the BH \citep{2011MNRAS.418L..79T}.
Signatures of magnetic flux eruptions have also been seen in the 3D pseudo-Newtonian magnetohydrodynamical (MHD) simulations presented by \cite{2008ApJ...677..317I} and \cite{2009ApJ...704.1065P}.

Such magnetic flux eruptions may have various interesting consequences, including ejection of magnetic flux tubes orbiting the BH (applied to observations with the GRAVITY instrument at the Very Large Telescope Interferometer of hotspots orbiting Sgr~A*, the supermassive black hole at the center of our Galaxy; \citealt{2020MNRAS.497.4999D,2021MNRAS.502.2023P}) and flares of energetic radiation \citep{2022MNRAS.511.3536S,2023ApJ...943L..29H,2024MNRAS.531.3961N}.
Eruptions should also be particularly relevant for the substructure and variability of BH crescent images obtained by the Event Horizon Telescope (EHT) Collaboration for M87* (the supermassive black hole at the center of M87) \citep{2019ApJ...875L...1E} and Sgr~A* \citep{2022ApJ...930L..12E}.
A much higher BH mass for M87*, $M_{\rm M87*} \simeq 6.5\times 10^9 M_\odot$ (with $M_\odot$ the solar mass) \citep{2019ApJ...875L...6E} implies a conveniently long dynamical time scale $GM_{\rm M87*}/c^3 \simeq 9\,{\rm h}$.
The image of M87* obtained by the EHT suggests the presence of a distinct bright hotspot in the ESE crescent sector \citep{2020A&A...634A..38N}, which also appears to be strongly depolarized \citep{2021ApJ...910L..12E}.
Moreover, during the 6-day long EHT campaign in April 2017, the image of M87* might have rotated counterclockwise by $\simeq 23^\circ$ \citep{2019ApJ...875L...4E}, which coincides \citep{2023mgm..conf..339N} with the rotation rate of the polarization vector for unresolved emission of M87* measured by ALMA \citep{2021ApJ...910L..14G} and corrected for Faraday rotation.
Although those results are considered not significant by the EHT Collaboration, future mm-VLBI (very long baseline interferometry) monitoring campaigns should be able to measure coherent rotation patterns of substructures in the M87* crescent image.
General relativistic magnetohydrodynamical (GRMHD) simulations predict a highly sub-Keplerian rotation rate of $\sim 1^\circ$ per $GM/c^3$ for ray-traced crescent image patterns, only very weakly depending on the BH spin or the sense of accretion flow angular momentum (prograde vs. retrograde) \citep{2023ApJ...951...46C}.

The driving mechanism of the BH magnetic flux eruptions had been unclear --- it has been interpreted as interchange modes \citep{2012MNRAS.423.3083M,2018MNRAS.478.1837M}, or as magneto-convection \citep{2022MNRAS.511.2040B}.
Recently, it was demonstrated by `extreme resolution' (effectively $5376 \times 2304 \times 2304$ cells) 3D ideal GRMHD simulations \citep{2022ApJ...924L..32R} to be the large-scale relativistic magnetic reconnection.
Extreme resolution allowed in particular to achieve thin current layers with sufficiently high Lundquist (magnetic Reynolds) number $S = v_{\rm A}L/\eta$ (with $v_{\rm A}$ the Alfv\'en speed, $L$ the current layer length, $\eta$ the magnetic diffusivity) that are unstable to tearing modes and break into chains of magnetic flux tubes (plasmoids) that determine a reconnection rate independent of the numerical magnetic diffusion.
These authors also compare their main results with a `low resolution' case ($288 \times 128 \times 128$ cells), providing several arguments against the reliability of the latter. In particular, reconnection rate based on the numerical diffusion is significantly overestimated.
On the other hand, strict adherence to conservation laws (energy, momentum, mass, magnetic flux) assures that low-resolution results are qualitatively robust, i.e., many features of magnetic flux saturation and eruptions can be reproduced (e.g., flux saturation levels, depth of density gaps, relativistic temperature levels, orbiting hotspot trajectories) even at a `test resolution' of $72 \times 64 \times 64$ \citep{2024hepr.conf}.

In this work we present the results of ideal GRMHD simulations of geometrically thick hot radiatively inefficient weakly magnetized accretion flows with a `standard' effective resolution of $288\times 256 \times 256$, similar to those presented in \cite{2022A&A...668A..66J}.
Our analysis is focused on investigating the initiation mechanism of magnetic flux eruptions, a problem that does not depend on the formation of plasmoid chains, but is centered on the formation of magnetic X-points.
In addition, we obtained qualitatively novel insights into the mechanism of regular accretion flows onto magnetically saturated BHs.

The 3D structure in quasi-spherical coordinates $(r,\theta,\phi)$ of accretion flows onto magnetically saturated BHs and magnetic flux eruptions cannot be captured by standard 2D maps in $(r,\theta)$ or $(r,\phi)$ coordinates.
Although the innermost accretion flow is constricted into a narrow disk, the disk is not strictly aligned with the equatorial plane or intrinsically symmetric.
For example, a slice along the equatorial plane may partially cross through a low-density magnetosphere bulging against a disk warp, which may be confused with a real density gap, or interpreted as signature of interchange instability \citep{2012MNRAS.423.3083M}.
Likewise, a latitudinal slice for constant $\phi$ cuts obliquely through the velocity and magnetic fields, making it very difficult to properly connect structures belonging to the same magnetic field line.
This problem is irreducibly 3D.

In order to obtain additional insight, we integrated large complete samples of magnetic field lines.
Those lines can be divided into two fundamental classes: connected or disconnected from the BH horizon (hereafter, the terms `connected' and `disconnected' shall mean just that; accordingly, `connection' of a line shall mean its transition from disconnected into connected, `disconnection' -- transition from connected into disconnected).
The disconnected lines are especially interesting, because they define a volume domain that in the magnetically saturated BH state is very flattened due to magnetic constriction, providing a fairly unobstructed view.
We use such line samples to divide the volume around an accreting BH into the connected and disconnected domains.
We identify the disconnected domain, which includes the entire accretion flow, winds and ejected magnetic flux tubes, as particularly useful to illustrate the initial development of magnetic flux eruptions.

In Section \ref{sec_methods}, we describe the setup of GRMHD numerical simulations and our method of field line integration.
Simulation results are presented in Section \ref{sec_results} according to their distributions in Kerr-Schild coordinates $(t,r,\theta,\phi)$,
with Section \ref{sec_res_t} comparing time histories,
Section \ref{sec_res_pro} presenting detailed results for our reference prograde simulation,
Section \ref{sec_res_retro} presenting selected results for the retrograde simulation,
Section \ref{sec_res_tphmaps} comparing $(t,\phi)$ diagrams of magnetic flux eruptions,
and Section \ref{sec_res_vperp} comparing $(r,\theta)$ maps of parameters measuring the alignment between velocity and magnetic fields.
Presentation of our reference simulation in Section \ref{sec_res_pro} includes:
maps in the $(r,\theta)$ coordinates (Section \ref{sec_res_pro_rthmaps});
maps in the $(r,\phi)$ coordinates (Section \ref{sec_res_pro_rphmaps});
3D structures of magnetic field lines
for regular accretion flow (Section \ref{sec_res_pro_lines_acc}),
for magnetic flux eruption (Section \ref{sec_res_pro_lines_erupt}),
and short-term time sequences (Section \ref{sec_res_pro_lines_seq}).
Section \ref{sec_disc} presents the discussion, and Section \ref{sec_conc} the conclusions.

We use natural Heaviside-Lorentz (HL) units with $c = G = 1$ and with the $4\pi$ factors absorbed into the definition of magnetic field ($B_{\rm HL}^i = B_{\rm G}^i/\sqrt{4\pi}$ with $B_{\rm G}^i$ in Gaussian units)\footnote{However, the dimensionless magnetic flux $\Phi$ is multiplied by $\sqrt{4\pi}$, as if based on $B_{\rm G}^i$.}.
The BH mass is also set to unity $M = 1$ and usually omitted.
Greek indices spanning $\{0,1,2,3\}$ are used for space-time forms (e.g., 4-vectors $X^\mu$), and Latin indices spanning $\{1,2,3\}$ are used for spatial forms (e.g., 3-vectors $X^i$).

\section{Methods}
\label{sec_methods}

\subsection{Setup of GRMHD simulations}
\label{sec_setup}

We performed 3D ideal GRMHD numerical simulations in the Kerr spacetime, using the public code {\tt Athena++} \citep{2020ApJS..249....4S,2016ApJS..225...22W}\footnote{\url{https://github.com/PrincetonUniversity/athena/wiki}}, which has been used to explore accretion flows on magnetically saturated BHs \citep{2019ApJ...874..168W}.
This code solves covariant equations of continuity $\nabla_\mu(\rho u^\mu) = 0$, energy-momentum conservation $\nabla_\mu T^{\mu\nu} = 0$, and Maxwell ${\nabla_\mu}^*\!F^{\mu\nu} = 0$.
In the above, $\nabla_\mu$ is the covariant derivative, $\rho$ is the plasma density, $u^\mu$ is the 4-velocity, $T^{\mu\nu} = (w+b^2)u^\mu u^\nu + (P + b^2/2)g^{\mu\nu} - b^\mu b^\nu$ is the energy-momentum tensor, $w = \rho + [\gamma/(\gamma-1)]P$ is the plasma enthalpy density, $\gamma$ is the adiabatic index, $P$ is the plasma pressure, $b^\mu$ is the magnetic 4-vector, $b^2$ is the magnetic enthalpy density ($b^2/2$ equals both the magnetic energy density and magnetic pressure), $g^{\mu\nu}$ is the inverse metric tensor, and $^*\!F^{\mu\nu} = b^\mu u^\nu - b^\nu u^\mu$ is the dual electromagnetic field tensor.
The magnetic 4-vector $b^\mu$ is related to the magnetic field 3-vector $B^i$ via $b^0 = g_{\mu i} u^\mu B^i$ (where $g_{\mu\nu}$ is the metric tensor) and $b^i = (B^i + u^i b^0)/u^0$, so that $^*\!F^{i0} = b^i u^0 - b^0 u^i = B^i$ \citep{2003ApJ...589..444G}\footnote{An alternative 3+1 spacetime splitting convention has $^*\!F^{i0} = B^i/\alpha$ with $\alpha$ the lapse \citep{2004MNRAS.350..427K}.}.
The code uses a constrained transport algorithm for cleaning the magnetic field divergence (ensuring that $\partial_i [\sqrt{-g}\,B^i] = 0$, where $g$ is the metric tensor determinant) \citep{2016ApJS..225...22W}.

The primitive parameters returned by the {\tt Athena++} code are plasma density $\rho$, plasma pressure $P$, 3 components of the fiducial velocity $\tilde{v}^i$, and 3 components of the magnetic field vector $B^i$.
The fiducial velocity is a projection $\tilde{v}^i = {\tilde{g}^i}_\mu u^\mu$ of the stationary 4-velocity $u^\mu = {\rm d}x^\mu/{\rm d}\tau$ (with $\tau$ the proper time; satisfying $g_{\mu\nu}u^\mu u^\nu = -1$) by the fiducial metric tensor $\tilde{g}_{\mu\nu} = g_{\mu\nu} + n_\mu n_\nu$, with $n_\mu = (-\alpha,0,0,0)$ the 4-velocity of the fiducial observer, and $\alpha = (-g^{00})^{-1/2}$ the lapse.
The stationary 4-velocity is calculated from $u^0 = (1 + g_{ij}\tilde{v}^i \tilde{v}^j)^{1/2}/\alpha$ and $u^i = \tilde{v}^i - \alpha^2 g^{0i} u^0$ \citep{2006ApJ...641..626N}.

Simulations were performed in the horizon-penetrating quasi-spherical Kerr-Schild coordinates $x^\mu = (t,r,\theta,\phi)$ for the spin of $|a| = 0.9$, for which the outer horizon radius is $r_{\rm H}/M = 1+\sqrt{1-a^2} \simeq 1.436$.
The metric tensor $g_{\mu\nu}$ non-zero components are:
$g_{tt} = -1+2r/\Sigma$,
$g_{rr} = 1+2r/\Sigma$,
$g_{\theta\theta} = \Sigma$,
$g_{\phi\phi} = (r^2+a^2+2ra^2\sin^2\theta/\Sigma)\sin^2\theta$,
$g_{tr} = g_{rt} = 2r/\Sigma$,
$g_{t\phi} = g_{\phi t} = -2ar\sin^2\theta/\Sigma$,
$g_{r\phi} = g_{\phi r} = -(1+2r/\Sigma)a\sin^2\theta$,
with $\Sigma = r^2 + a^2\cos^2\theta$;
the determinant of this metric is $g = -\Sigma^2\sin^2\theta$.

Our coordinate grid is uniform in $\phi$, $\theta$ and $\log{r}$.
We used static mesh refinement (SMR) in order to reduce resolution around the coordinate axis ($\theta \simeq 0,\pi$).
In our simulations, we used 3 levels of SMR, with the Level-3 grid of $N_{r,\rm L3} \times N_{\theta,\rm L3} \times N_{\phi,L3} = 144 \times 120 \times 256$ cells covering the region of $r_{\rm L3,min} \le r \le r_{\rm L3,max}$, $\theta_{\rm L3,min} = (7/24)\pi = 52.5^\circ \le \theta \le \theta_{\rm L3,max} = (17/24)\pi = 127.5^\circ$, $0 \le \phi/\pi < 2$.
Radial vertices of the Level-3 grid are defined as $r_{\rm f}[i] = r_{\rm H} (r_{\rm max}/r_{\rm H})^{(i-N_{\rm H})/(2N_{r,\rm L3}-N_{\rm H})}$ for $i = 0,1,...,N_{r,\rm L3}$, where $N_{\rm H} = 8$ is the number of Level-3 cells inside the horizon\footnote{The Level-0 grid has exactly 1 cell inside the horizon, in which we followed \cite{2019ApJ...874..168W} -- this is limited to the polar regions ($\theta < 30^\circ$ and $\theta > 150^\circ$) entirely within the magnetosphere; we performed full-resolution tests with up to 8 Level-0 cells inside the horizon, the results of which show no significant differences outside the horizon.}, and $r_{\rm max} = 500 M$ is the maximum radius of the Level-0 grid; hence, $r_{\rm L3,min} = r_{\rm f}[0] \simeq 1.21 M$ and $r_{\rm L3,max} = r_{\rm f}[N_{r,\rm L3}] \simeq 24.6 M$.

Our simulations were initiated from an axisymmetric untilted torus in hydrodynamical equilibrium described by \cite{1976ApJ...207..962F}.
For $|a| = 0.9$, we considered the prograde case (effective spin of $a = 0.9$) and the retrograde case (effective spin of $a = -0.9$).
The torus extends from an inner radius $r_{\rm in}$ ($6 M$ for $a = 0.9$; $11.25 M$ for $a = -0.9$) to a fixed outer edge $r_{\rm out} = 70M$ (the specific enthalpy satisfies $h(6M) = h(r_{\rm in}) = h(r_{\rm out})$; in the prograde case $r_{\rm in} = 6M$; in the retrograde case $r_{\rm in} > 6M$ is another root of $h$).
The density peak $\rho_{\rm peak} = 1$ (in code units) and the pressure peak $P_{\rm peak}$ ($0.0066$ for $a = 0.9$; $0.0026$ for $a = -0.9$) are located at the same radius $r_{\rm peak}$ ($13 M$ for $a = 0.9$; $22.2 M$ for $a = -0.9$), at which we evaluate the Keplerian specific angular momentum $l_{\rm peak} = l(r_{\rm peak})$, which is assumed to be constant across the torus.
The magnetic vector potential was set as $A_\phi \propto r^5\,\max\{\rho-\rho_{\rm max}/20,0\}$, peaking at the amplitude of $2\times 10^{-7}$ in code units at the radius $r_{\rm peak,A}$ ($34 M$ for $a = 0.9$; $44 M$ for $a = -0.9$), so that at $r_{\rm peak}$ the plasma parameter is $\beta_{\rm peak} = 2P_{\rm peak}/b_{\rm peak}^2$ ($1920$ for $a = 0.9$; $1150$ for $a = -0.9$) with $b_{\rm peak}^2 \equiv b^2(r_{\rm peak})$.
We also introduced within the initial torus a random perturbation to the radial component of fiducial 4-velocity $\tilde{u}^r$ with amplitude of $\pm 5\%$.

We also set numerical floor profiles of density $\rho_{\rm floor}(r) = \max\{10^{-4}r^{-1.5},10^{-6}\}$ and pressure $P_{\rm floor}(r) = \max\{10^{-6}r^{-2.5},10^{-8}\}$, and limits on the plasma parameter $\beta_{\rm min} = 10^{-3}$, magnetization $\sigma_{\rm max} = 100$ and bulk Lorentz factor $u^t_{\rm max} = 50$.
The adiabatic index was set as $\gamma = 13/9$ \citep[e.g.,][]{2019ApJ...874..168W}.

In the configuration of the {\tt Athena++} code, we selected the Harten-Lax-van~Leer-Einfeldt (HLLE) Riemann solver, and the piecewise parabolic method (PPM) for the spatial reconstruction of primitive variables \citep{2019ApJ...874..168W,2019ApJS..243...26P}.
For time integration, we selected the van Leer predictor-corrector scheme for long-term runs, and the 4-th order Runge-Kutta scheme for short-term runs with frequent output of full data cubes, in both cases setting the Courant-Friedrichs-Lewy parameter ${\rm CFL} = 0.5$.

\subsection{Magnetic field line integration}

Magnetic field lines have been integrated in the Kerr-Schild coordinates $(r,\theta,\phi)$ (with $r$ in units of $M$ and $\theta,\phi$ in radians) using a 4th-order Runge-Kutta scheme.
The input consists of data cubes for components of magnetic field 3-vector $B^r_{n1,n2,n3},B^\theta_{n1,n2,n3},B^\phi_{n1,n2,n3}$ at grid nodes $(r_{n1},\theta_{n2},\phi_{n3})$.
For every position of interest, local field components were calculated by trilinear interpolation using the 8 grid nodes at the vertices of the volume cell including that position.
Position $x_a^i$ would be advanced to $x_{a+1}^i = x_a^i + (B^i/|B|)\,\delta l$, using a simplified norm $|B|^2 = \sum_i(B^i)^2$ (no difference was found from using the metric norm) and fixed integration step $\delta l = 0.02$.
We tested for accuracy against simple analytical solutions, and for convergence in $\delta l$.
Integration was performed within the Level-3 SMR region and over the radial range $r_{\rm H} < r < r_{\rm max}$, with $r_{\rm max} = 6 M$.
Line samples were seeded from fixed positions at $r = r_0$ ($r_0 = 5.9 M$ for disconnected lines; $r_0 = 1.42 M$ for doubly connected lines) and for every grid element in $\theta_{\rm L3,min} < \theta_0 < \theta_{\rm L3,max}$ and $0 \le \phi_0 < 2\pi$.

\begin{figure}
\includegraphics[width=\columnwidth,bb=0 0 432 288]{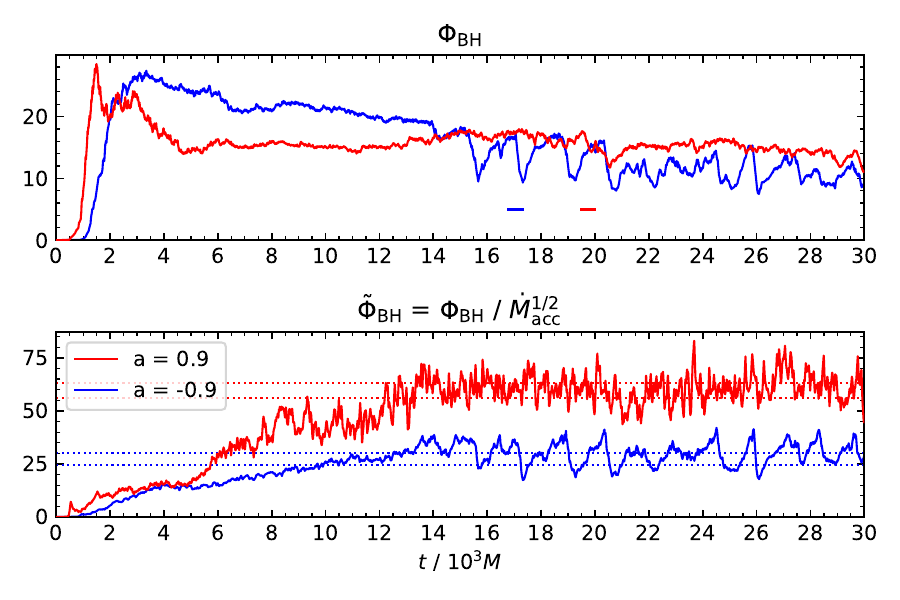}
\caption{Long-term histories of the BH magnetic flux $\Phi_{\rm BH}$ (upper panel) and of its value normalized by the mass accretion rate $\tilde\Phi_{\rm BH} = \Phi_{\rm BH}/\dot{M}_{\rm acc}^{1/2}$ (lower panel) for 2 values of effective spin $a$.
Short bars in the upper panel indicate the time windows covered by Figure \ref{fig_hist_short}.}
\label{fig_hist_long}
\end{figure}

\begin{figure}
\includegraphics[width=\columnwidth,bb=0 0 432 252]{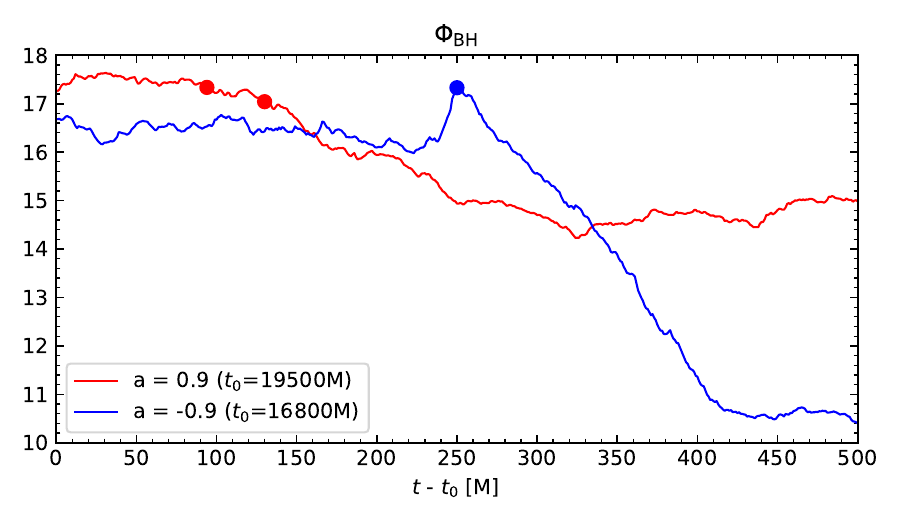}
\caption{Short-term histories of the BH magnetic flux $\Phi_{\rm BH}$ during magnetic flux eruptions in the prograde $a = 0.9$ (red; time window from $t_0 = 19500 M$ to $20000 M$) and retrograde $a = -0.9$ (blue; time window from $t_0 = 16800 M$ to $17300 M$) cases.
Two epochs of interest are indicated in the prograde case: the eruption onset at $t = 19594 M$, and its advanced stage at $t = 19630 M$;
one epoch is indicated in the retrograde case: the magnetic flux peak at $t = 17050 M$.}
\label{fig_hist_short}
\end{figure}

\begin{figure*}
\includegraphics[width=\textwidth,bb=0 0 1166 360]{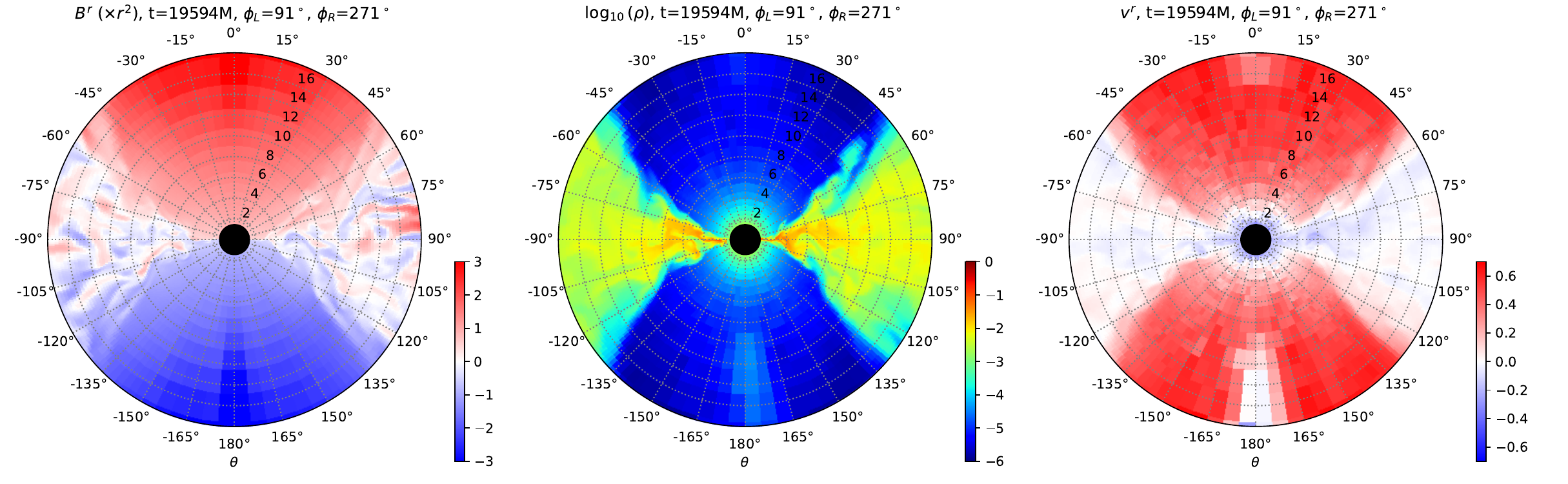}
\caption{Overview maps ($r \le 18 M$) in the Kerr-Schild $(r,\theta)$ coordinates of selected parameters for the prograde $a = 0.9$ case at $t = 19594 M$.
From the left, the first panel shows the radial component of magnetic field 3-vector $B^r$ (multiplied by $r^2$), the second panel shows the plasma density $\rho$ in log scale, and the third panel shows the radial component of velocity 3-vector $v^r$.
The left and right halves of every panel show slices for opposite azimuths $\phi_{\rm L},\phi_{\rm R}$ indicated in the titles.
The black circles mark the BH with horizon radius $r_{\rm H} \simeq 1.436 M$.
The coarse cells in the polar regions ($|\theta| < 30^\circ$ and $|\theta| > 150^\circ$) result from the SMR.}
\label{fig_rthmaps_rmax18}
\end{figure*}

\begin{figure*}
\begin{center}
\includegraphics[width=\graphreduce\textwidth,bb=0 0 1080 360]{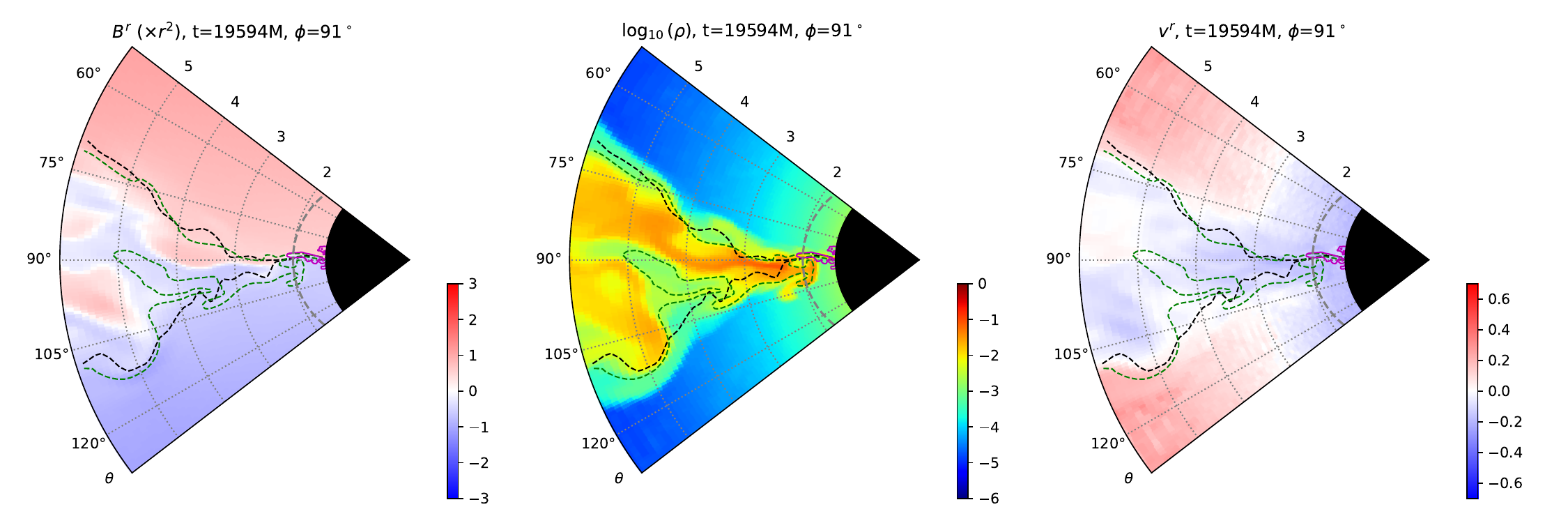}
\includegraphics[width=\graphreduce\textwidth,bb=0 0 1080 360]{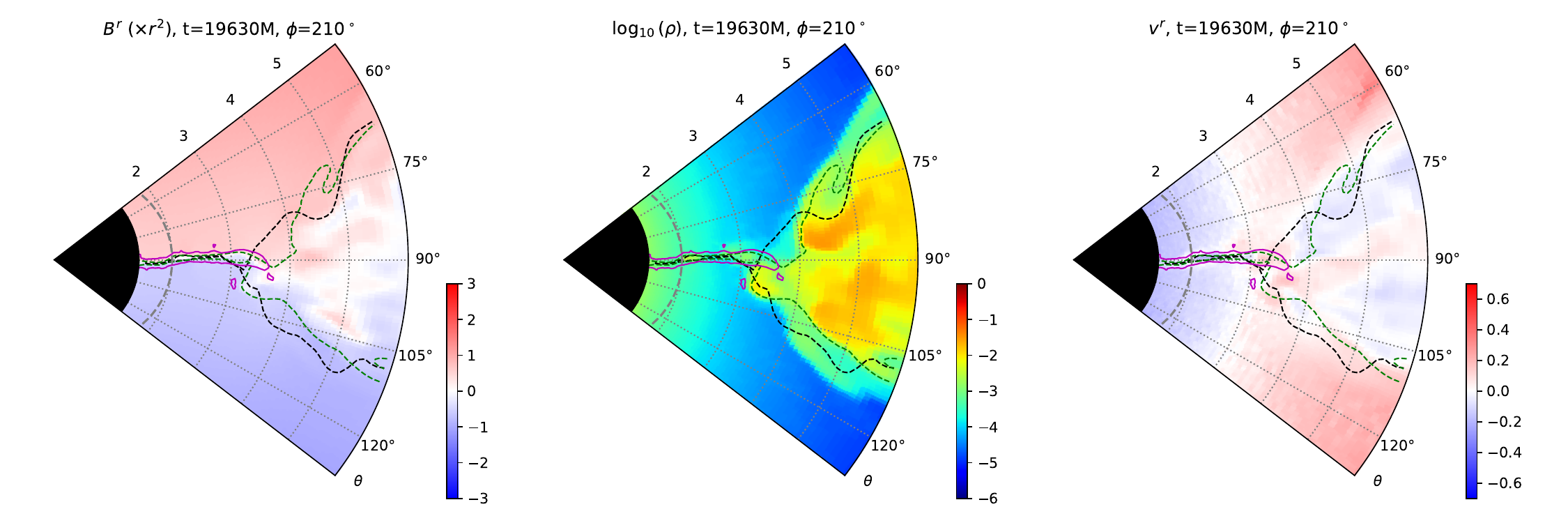}
\end{center}
\caption{Close-up maps ($r \le 6 M$) in the $(r,\theta)$ coordinates (with $\theta$ limited to the Level-3 SMR domain) of the same parameters as in Figure \ref{fig_rthmaps_rmax18} (from left to right) for the prograde $a = 0.9$ case at two epochs: $t = 19594 M$ (upper row of panels) and $t = 19630 M$ (lower row of panels).
Each panel shows a slice for the azimuth $\phi$ indicated in its title (also marked in Figure \ref{fig_rphmaps}).
The gray dashed circles mark the ergosphere boundary $r_{\rm E}(\theta)$, the black dashed lines mark the boundary of domain magnetically disconnected from the BH horizon, the green dashed lines mark equipartition between magnetic and plasma enthalpies ($\sigma = b^2/w = 1$),
the magenta contours mark relativistic temperature $\log_{10} T = 0.5$.}
\label{fig_rthmaps_rmax6}
\end{figure*}

\section{Results}
\label{sec_results}

\subsection{Histories in $(t)$}
\label{sec_res_t}

Figure \ref{fig_hist_long} presents long-term histories of the BH magnetic flux $\Phi_{\rm BH}$ and of its value normalized by the mass accretion rate $\tilde\Phi_{\rm BH} \equiv \Phi_{\rm BH}/\dot{M}_{\rm acc}^{1/2}$.

In the prograde $a = 0.9$ case, the BH magnetic flux peaks at $\Phi_{\rm BH,peak} \simeq 28$ at $t \simeq 1500 M$, then settles to $\Phi_{\rm BH} \sim 15$ by $t \simeq 5000 M$, and then remains roughly constant until at least $t \sim 30000 M$.
The normalized flux $\tilde\Phi_{\rm BH}$ builds up much more slowly, achieving saturation at the level of $\tilde\Phi_{\rm BH} \sim 65$ by $t \sim 14000 M$.
Relatively shallow magnetic flux eruptions can be seen from $t \sim 18000 M$.

In the retrograde $a = -0.9$ case, the BH magnetic flux peaks at the $\Phi_{\rm BH,peak} \simeq 27$ at $t \simeq 3300 M$, followed by a slow decay.
The normalized flux $\tilde\Phi_{\rm BH}$ achieves saturation at the level of $\tilde\Phi_{\rm BH} \sim 30$ by $t \sim 14000 M$\footnote{Our flux saturation levels are slightly higher than the results of longer simulations of comparable resolution reported in \cite{2022MNRAS.511.3795N} -- $\tilde\Phi_{\rm BH} \simeq 56$ for $a = 0.9$ and $\tilde\Phi_{\rm BH} \simeq 25$ for $a = -0.9$.}.
Clear magnetic eruptions, deeper than in the prograde case, appear regularly after $t \sim 15000 M$.

For detailed analysis, we selected the magnetic flux eruption beginning around $t \sim 19500 M$ in the prograde $a = 0.9$ case.
The short-term history of $\Phi_{\rm BH}$ during that eruption is shown in Figure \ref{fig_hist_short}.
During this eruption, the BH magnetic flux decreases from $\Phi_{\rm BH} \simeq 17.6$ at $t \simeq 19530 M$ to $\simeq 14.2$ at $t \simeq 19825 M$, a decrease of $\simeq 19\%$ over $\Delta t \simeq 300 M$.
The flux decrease is not exactly uniform and monotonic.
Further analysis will be focused on two moments of time: at $t = 19594 M$, which has been identified as the actual onset of the eruption, and at $t = 19630 M$, in early advanced stage of the eruption.

In the retrograde $a = -0.9$ case, we will focus on the moment of peak $\Phi_{\rm BH} \simeq 17.3$ at $t = 17050 M$. This is followed by a decrease to $\Phi_{\rm BH} \simeq 10.7$ ($\simeq 38\%$) over $\Delta t \simeq 165 M$, but it is preceded by rapid increase of $\Phi_{\rm BH}$ by $\simeq 6.5\%$.

\subsection{Reference prograde simulation}
\label{sec_res_pro}

\subsubsection{Maps in $(r,\theta)$ for slices in $\phi$}
\label{sec_res_pro_rthmaps}

Figure \ref{fig_rthmaps_rmax18} shows overview maps in the $(r,\theta)$ coordinates within $r \le 18 M$ for the prograde $a = 0.9$ case at $t = 19594 M$.
Two main regions can be distinguished:
(1) geometrically thick (but converging to a thin nozzle near the BH) accretion flow centered around the equatorial plane $\theta = \pm 90^\circ$, characterized by negative radial velocity $v^r < 0$, high plasma density ($\log_{10}\rho > -3$) and turbulent radial magnetic field $B^r$ (with multiple current layers);
(2) paraboloidal (hourglass) bipolar magnetosphere centered around the polar axis $\theta = 0,180^\circ$, characterized by positive radial velocity $v^r > 0$, low plasma density ($\log_{10}\rho < -3$; mostly at the floor level) and ordered split-monopole radial magnetic field $B^r$.
Very strong density contrast means that the magnetosphere is relativistically magnetized ($\sigma = b^2/w > 1$), with radial outflows accelerating over distance to relativistic speeds, destined to become collimated into relativistic jets.

Figure \ref{fig_rthmaps_rmax6} presents close-up $(r,\theta)$ maps ($r \le 6 M$) of the same parameters ($B^r\,(\times r^2)$, $\log_{10}\rho$ and $v^r$).
For $t = 19594 M$, we show the $\phi \simeq 90^\circ$ slices (corresponding to the left halves of each panel in Figure \ref{fig_rthmaps_rmax18}) with a mostly regular accretion flow, relatively thick, dense and cold.
Nevertheless, the plasma density $\rho$ distribution is significantly perturbed, not symmetric about the equatorial plane.
The green contours of enthalpy equipartition ($\sigma = 1$) closely follow the density distribution.
The boundary between magnetically connected and disconnected domains is roughly close to the $\sigma = 1$ lines with significant local departures.
Hence, there exist $\sigma > 1$ regions magnetically disconnected from the horizon, and $\sigma < 1$ regions magnetically connected to the horizon.
The boundary between positive and negative $B^r$ domains lies close to the equatorial plane, roughly middle through the accretion flow; it is sharp, indicating a strong equatorial current layer.
There are no signs of magnetic discontinuities (current layers) between the connected and disconnected domains.
Similar structures can also be seen in $(r,\theta)$ maps of the $B^\phi$ component, which is of comparable strength to $B^r$.
The $B^\theta$ component is weaker by order of magnitude, therefore, disconnected field lines within $r < 6M$ can be expected for form spirals with $|B^r| \sim |B^\phi| \gg |B^\theta|$.

For $t = 19630 M$, we present a different slice at $\phi \simeq 210^\circ$ across an ongoing magnetic flux eruption.
Here, the accretion flow extends only for $r > 3.5 M$, showing a characteristic inner tip that we will argue to be a magnetically arrested chokepoint.
However, the current layer extends all the way to the horizon.
Its low-density inner section is immersed in a relatively thin layer of relativistic temperature ($\log_{10}T > 0.5$), which was limited to $r < 2 M$ at $t = 19594 M$.
The boundary between magnetically connected and disconnected domains is broader than the equipartition line for $3.5 M < r < 4 M$, which favors the presence of $\sigma > 1$ disconnected lines.

\begin{figure*}[t]
\begin{center}
\includegraphics[width=\graphreduce\textwidth,bb=0 0 1166 360]{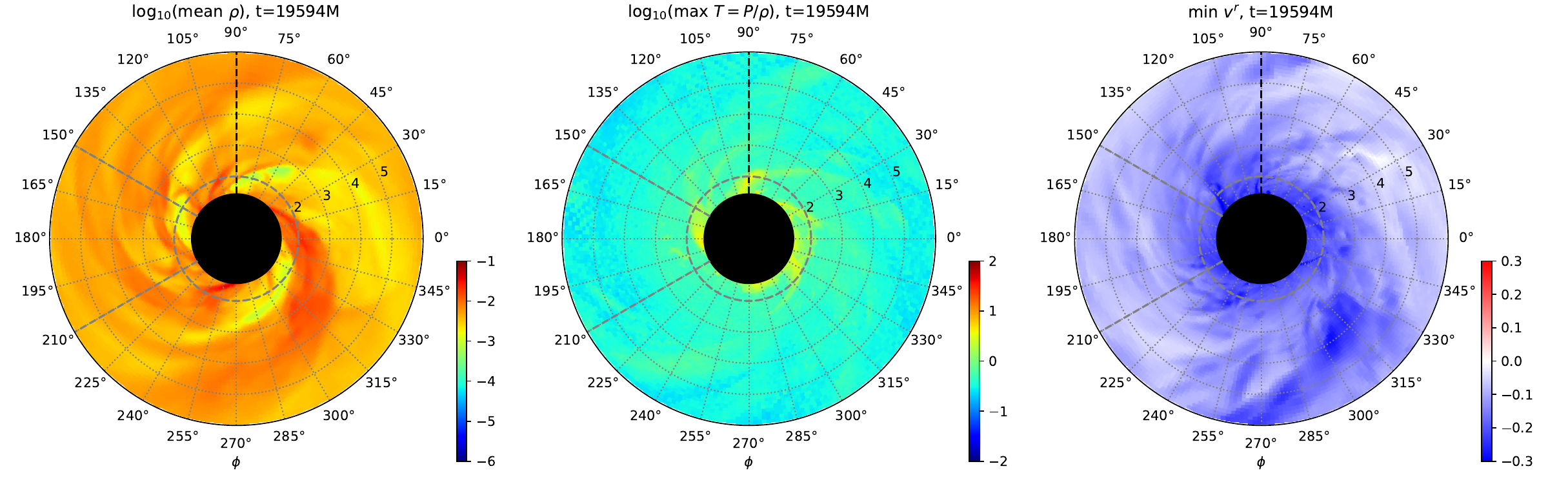}
\includegraphics[width=\graphreduce\textwidth,bb=0 0 1166 360]{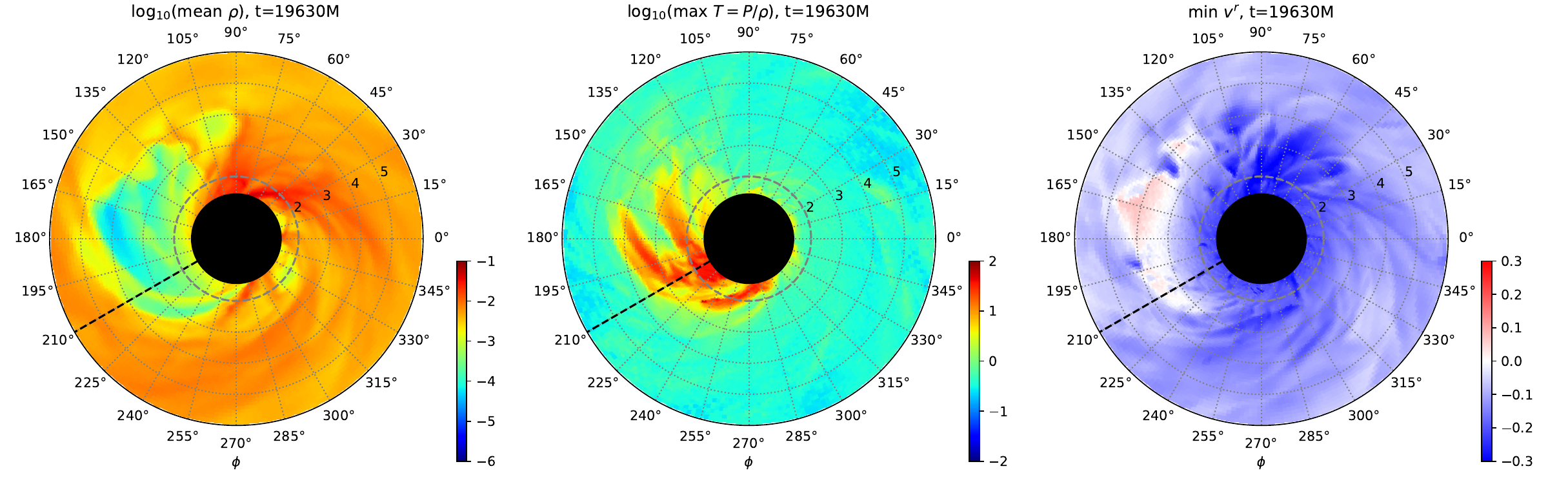}
\end{center}
\caption{Maps in the $(r,\phi)$ coordinates of parameter statistics drawn from latitudinal arcs $\theta_{\rm L3,min} < \theta < \theta_{\rm L3,max}$ for the $a = 0.9$ case at two epochs: $t = 19594 M$ (upper row of panels) and $t = 19630 M$ (lower row of panels).
From the left, the panels show: (1) mean density $\rho_{\rm mean}$ in log scale, (2) maximum temperature $T_{\rm max}$ in log scale, (3) minimum radial velocity $v^r_{\rm min}$.
The black circles mark the BH with horizon radius $r_{\rm H} \simeq 1.436 M$,
the gray dashed circles mark the $\theta = 90^\circ$ ergosphere boundary $r_{\rm E} = 2 M$,
the gray dashed lines in the upper panels mark the cutout $\phi$ sector in Figures \ref{fig_lines_a09_onset_sigma} and \ref{fig_lines_a09_onset_vr},
and the black dashed lines mark the $\phi$ values for the $(r,\theta)$ maps presented in Figure \ref{fig_rthmaps_rmax6}.}
\label{fig_rphmaps}
\end{figure*}

\subsubsection{Maps in $(r,\phi)$ for statistics over $\theta$}
\label{sec_res_pro_rphmaps}

In Figure \ref{fig_rphmaps}, we present close-up maps ($r \le 6 M$) in the $(r,\phi)$ coordinates of parameters: plasma density $\rho$, plasma temperature $T = P/\rho$ (with $P$ the plasma pressure), and radial component of the velocity 3-vector $v^r$.
Recognizing from Figure \ref{fig_rthmaps_rmax6} that the innermost accretion flow is constricted to a thin disk that can be locally warped and azimuthally modulated, we do not simply slice the data cubes along the equatorial plane ($\theta = 90^\circ$).
Instead, for each value of $(r,\phi)$ we calculate a parameter statistic over the range of $\theta_{\rm L3,min} < \theta < \theta_{\rm L3,max}$.
For the plasma density, we take the mean value $\rho_{\rm mean}$, which is dominated by the densest part of the accretion flow, but is also partially affected by the local flow thickness, and only in the gap regions it approaches the much lower floor values.
For the plasma temperature, we take the maximum value $T_{\rm max}$, which captures even very thin active current layers with relativistic heating due to magnetic reconnection.
For the radial velocity, we take the minimum value $v^r_{\rm min}$, which captures the regions in which radial outflow spans the entire range of $\theta$, i.e., regions of completely reversed accretion.

At $t = 19594 M$, at the onset of magnetic flux eruption, almost all $\phi$ sectors show high density ($\log_{10}\rho_{\rm mean} > -3$), non-relativistic temperature ($\log_{10}T_{\rm max} < 0$) and radial inflows ($v^r_{\rm min} < -0.1$), i.e., a regular accretion flow that in principle can advect magnetic flux inwards, supporting flux accumulation at the BH horizon (except that $\Phi_{\rm BH}$ is saturated).
The line of $\phi = 90^\circ$, corresponding to the slices shown in the upper row of panels in Figure \ref{fig_rthmaps_rmax6}, crosses through regions that are largely dense, cold and accreting, except for a narrow spiral density gap crossing that line for $r < 2 M$, and a minor temperature enhancement at $r \simeq 2 M$ that will be identified as the actual onset of magnetic flux eruption.

At $t = 19630 M$, a large region of relativistic temperature can be seen, touching the BH horizon over a range of $180^\circ < \phi < 300^\circ$, and extending to $r \simeq 4 M$ at $\phi \simeq 180^\circ$.
This coincides with a deep density gap ($\log_{10}\rho_{\rm mean} \simeq -4$), and with the presence of radial outflows for $r > 3.5 M$ along the spiral extension of the high-temperature patch.
The line of $\phi = 210^\circ$, corresponding to the slices shown in the lower row of panels in Figure \ref{fig_rthmaps_rmax6}, crosses through that eruption for $r < 3.5 M$, touching the region of completely reversed accretion at $r \simeq 3.5 M$.

\subsubsection{Magnetic field lines: regular accretion flow}
\label{sec_res_pro_lines_acc}

We now present a large sample of magnetic field lines in the prograde $a = 0.9$ case at $t = 19594 M$.
In order to make the sample statistically complete, we seeded individual lines from a grid of fixed positions at $r_0 = 5.9 M$ for every grid element in $\theta_{\rm L3,min} < \theta_0 < \theta_{\rm L3,max}$ and $0 < \phi_0 < 2\pi$.
Depending on its minimum radius $r_{\rm min}$, each line was classified as connected ($r_{\rm min} \le r_{\rm H}$) or disconnected ($r_{\rm min} > r_{\rm H}$).
We also present a complete sample of doubly connected field lines, seeded just under the horizon at $r_0 = 1.42 M$ and reaching a maximum radius satisfying $r_{\rm H} < r_{\rm max} < 6M$.

\begin{figure}
\includegraphics[width=\columnwidth,bb=0 0 1600 1170]{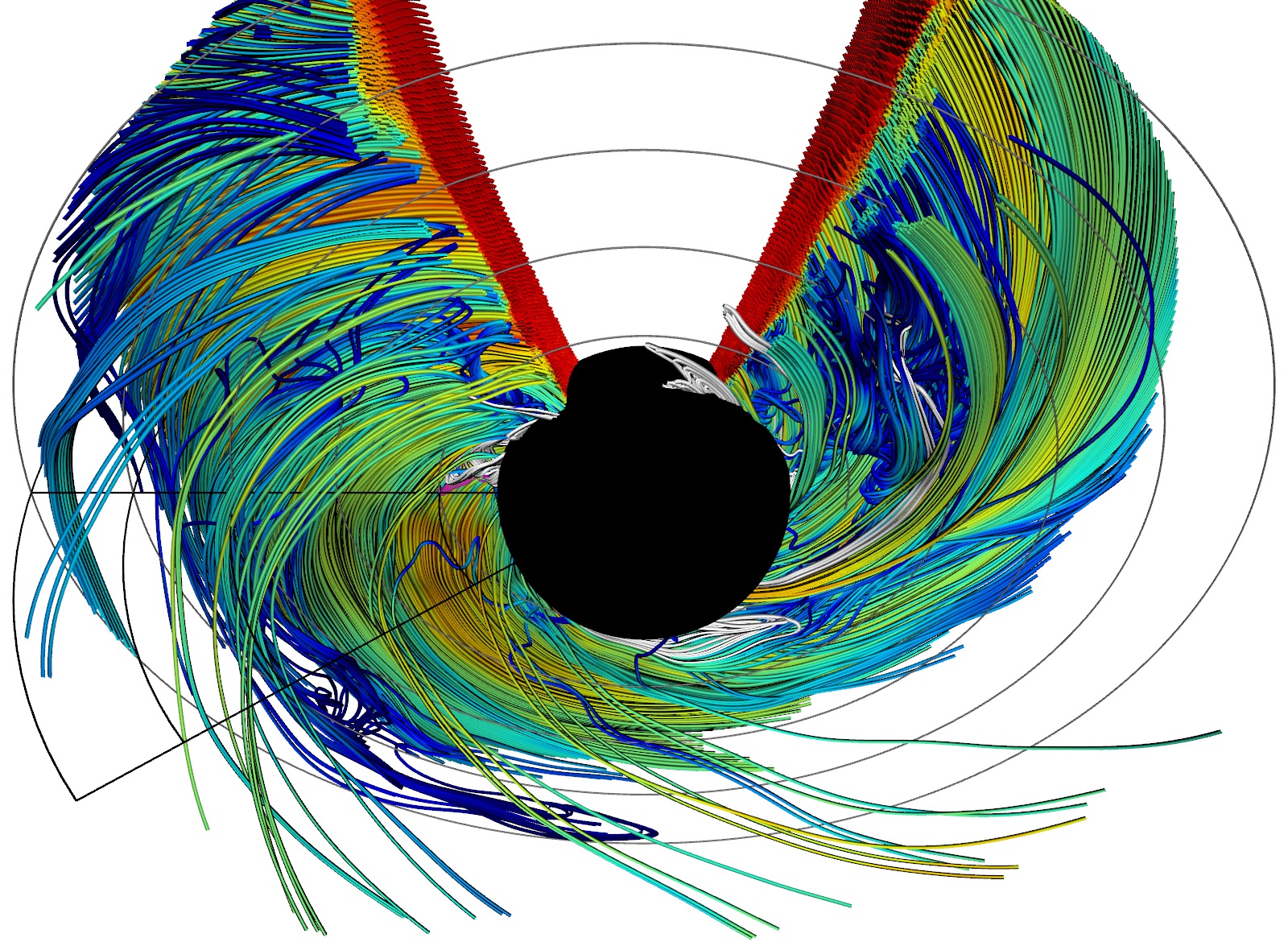}
\includegraphics[width=\columnwidth,bb=0 0 1600 1180]{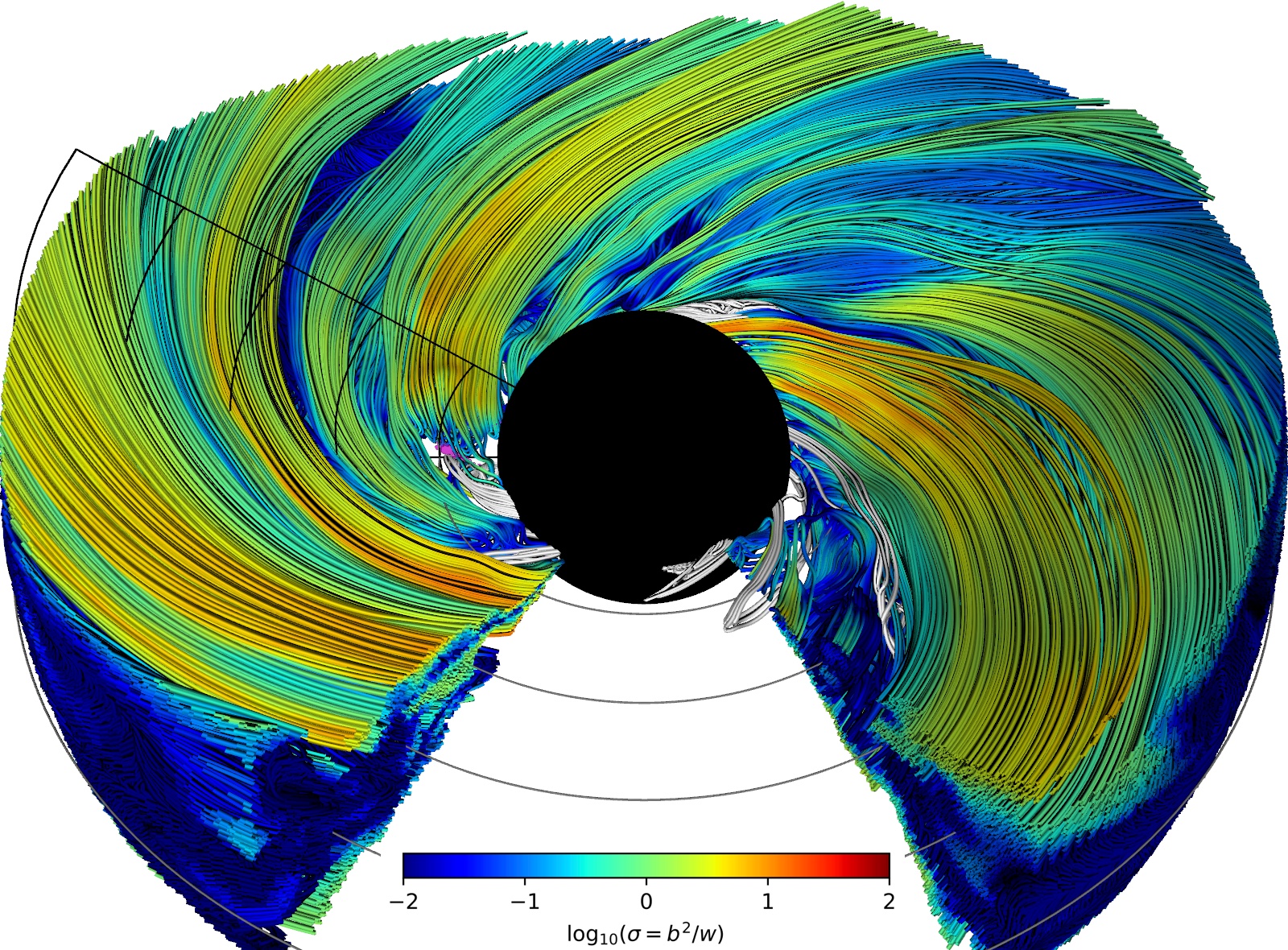}
\caption{Selected magnetic field lines within $r \le 6 M$ in the prograde $a = 0.9$ case at the onset of magnetic flux eruption ($t = 19594 M$), colored by the magnetization parameter $\sigma = b^2/w$.
Note that line colors are also affected by shading, with lines seen at small angles rendered darker.
Upper panel: horizon-connected lines seeded at $r_0 = 5.9 M$ for $\theta_{\rm L3,min} < \theta_0 < 90^\circ$, seen from the latitude $\theta_{\rm obs} = 135^\circ$;
lower panel: horizon-disconnected lines seeded at $r_0 = 5.9 M$ for $\theta_{\rm L3,min} < \theta_0 < \theta_{\rm L3,max}$, seen from the latitude $\theta_{\rm obs} = 45^\circ$.
In both panels, a small but $(\theta_0,\phi_0)$-complete sample of doubly connected lines (closed loops with $\max(r) < 6 M$) seeded at $r_0 = 1.42 M$ (just under the horizon) are shown colored in white.
Both images are seen from the same azimuth $\phi_{\rm obs} = 180^\circ$; data for $150^\circ < \phi < 210^\circ$ (except for the doubly connected lines) are cut out to reveal $(r,\theta)$ sections.
Black spheres indicate the outer horizon at $r_{\rm H} = 1.436 M$.
The grid of arcs marks the radii $r/M = 2,3,4,5,6$ in the equatorial plane $\theta = 90^\circ$ (gray), and in the $\phi = 90^\circ$ quadrant spanning the latitudes $\theta_{\rm L3,min} \le \theta \le \theta_{\rm L3,max}$ (black).
A small magenta patch just to the left from the horizon marks relativistic temperature ($\log_{10}T \ge 0.5$).
Associated movie is available online.}
\label{fig_lines_a09_onset_sigma}
\end{figure}

Figure \ref{fig_lines_a09_onset_sigma} presents separately the samples of connected lines for $\theta_0 < 90^\circ$ (upper panel), and all disconnected lines (lower panel), colored by the magnetization $\sigma = b^2/w$.
The distribution of connected lines appears somewhat untidy -- a small number of individual connected lines can be seen sticking out in the foreground, they seem to navigate narrow channels within the disconnected domain.
The disconnected lines appear more tidy -- groups of neighboring lines form coherent flux tubes.
The doubly connected lines form several short spiral loops. These are coherent but short-lived structures, collapsing towards the horizon along with the local accretion flow, slightly accelerated by the line tension.
Additional animations show that they appear to form rather spontaneously.

Most of the disconnected lines are dominated by the plasma enthalpy ($w \gg b^2$; blue), and most of the connected lines are dominated by the magnetic enthalpy ($b^2 \gg w$; red).
Such distribution of $\sigma$ reflects a strong latitudinal (along $\theta$) stratification of the plasma density $\rho$, which can be seen in the $(r,\theta)$ maps (Figures \ref{fig_rthmaps_rmax18} and \ref{fig_rthmaps_rmax6}).
On the other hand, the magnetic enthalpy density $b^2$ is roughly uniform in $\phi$ and $\theta$ (except for the equatorial current layer), and variations in the temperature $T = P/\rho$ (affecting the relation between $w$ and $\rho$) are less prominent.

The interface between the disconnected and connected domains is dominated by lines roughly in enthalpy equipartition ($w \sim b^2$; green).
However, exceptions can be seen in the form of tubes of highly magnetized disconnected lines.
Note that magnetization $\sigma$ (like plasma density $\rho$) is fairly uniform along the field lines, but stratified across them.
This reflects the ideal MHD principle that plasma is free to move along the magnetic field, but motions across the magnetic field are restricted, as they should be equivalent to transporting the field.
Density redistribution is straightforward along individual field lines, but cross-field diffusion (motion between neighboring field lines) is possible only in regions where ideal MHD is violated due to numerical limitations.

One high-$\sigma$ magnetic flux tube approaches from behind the BH to a small magenta patch of relativistic temperature that can be seen in both panels across the $\phi = 90^\circ$ quadrant at $r < 2 M$ and $\theta \simeq 90^\circ$.
One can notice a corresponding gap of matching position and shape in $(r,\phi)$ in the sample of connected lines (upper panel of Figure \ref{fig_lines_a09_onset_sigma}), where moderately magnetized lines ($\sigma \sim 1$; green) are missing.
Around this relativistically hot patch, disconnected field lines meet (and partially interlock) with a slightly twisted group of doubly connected lines.
In Section \ref{sec_res_pro_lines_seq} we shall demonstrate that those field lines are interacting dynamically.

\begin{figure}
\includegraphics[width=\columnwidth,bb=0 0 1600 1180]{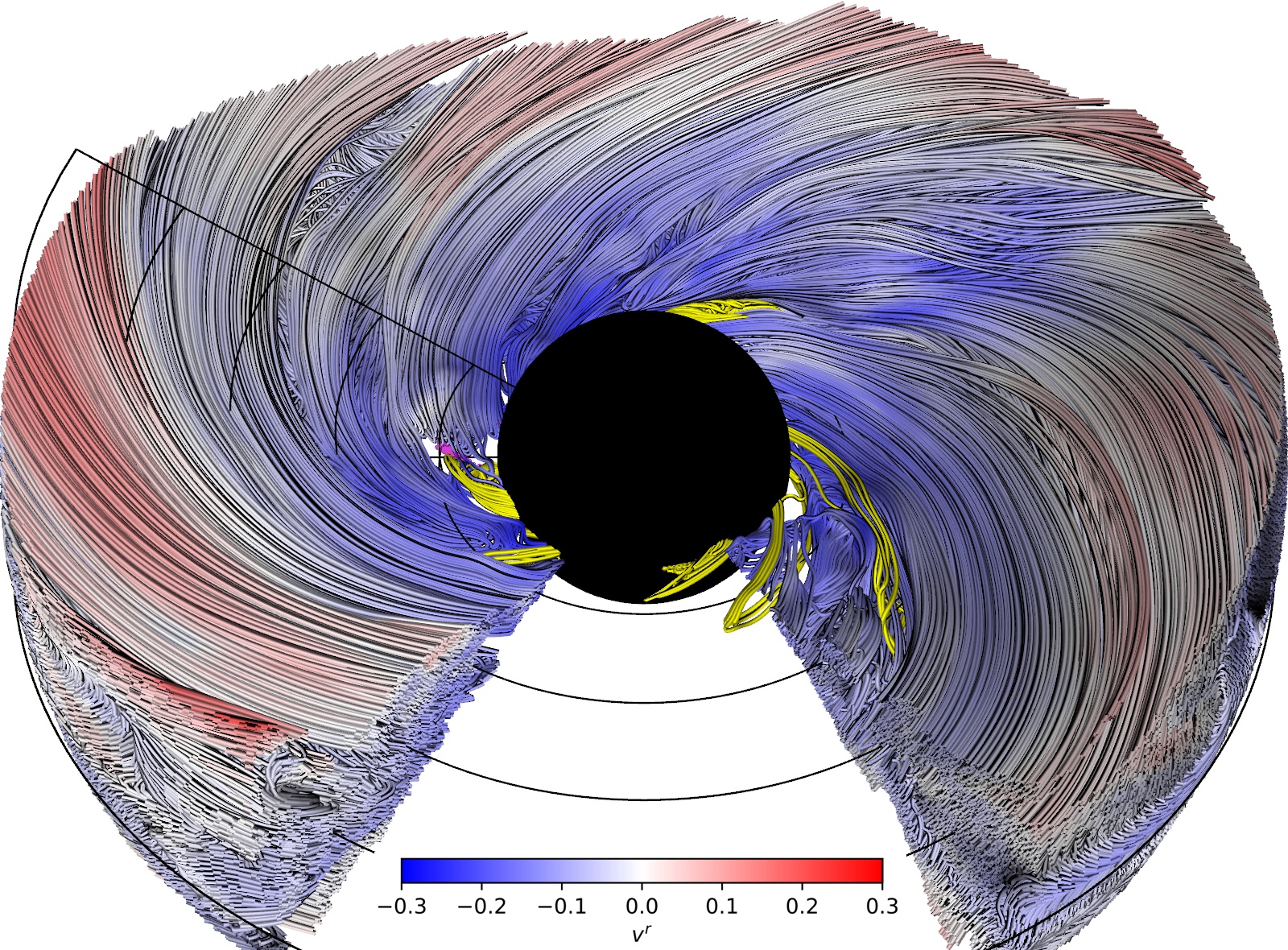}
\caption{Same as the lower panel of Figure \ref{fig_lines_a09_onset_sigma}, but with the horizon-disconnected lines colored by the radial 3-velocity $v^r$, and doubly-connected lines colored yellow.}
\label{fig_lines_a09_onset_vr}
\end{figure}

Figure \ref{fig_lines_a09_onset_vr} presents the same disconnected lines colored by the radial component of plasma 3-velocity $v^r = u^r / u^t$.
Blue regions indicate $v^r < 0$, i.e., radial inflow or accretion.
The disconnected domain is dominated by fast accretion with $v^r$ up to 0.3, especially along the equatorial plane.
However, a small part of those field lines, especially those elevated furthest away from the equatorial plane in their outer sections ($r > 4 M$), show $v^r > 0$ (red), i.e., radial outflows.
Those lines form a wind connected to the accretion flow often in its innermost regions ($r < 2 M$).

The distributions of $\sigma$ and $v^r$ along individual disconnected and elevated field lines does not appear to be highly correlated.
The most mass-depleted disconnected lines (orange on the $\log\sigma$ color scale) might be expected to form an outflowing wind.
Most of the elevated disconnected lines show stagnation points (white on the $v^r$ color scale) and an outflow (red) within $r < 6M$, regardless of their magnetization.

\subsubsection{Magnetic field lines: magnetic flux eruption}
\label{sec_res_pro_lines_erupt}

\begin{figure}
\includegraphics[width=\columnwidth,bb=0 0 1600 1224]{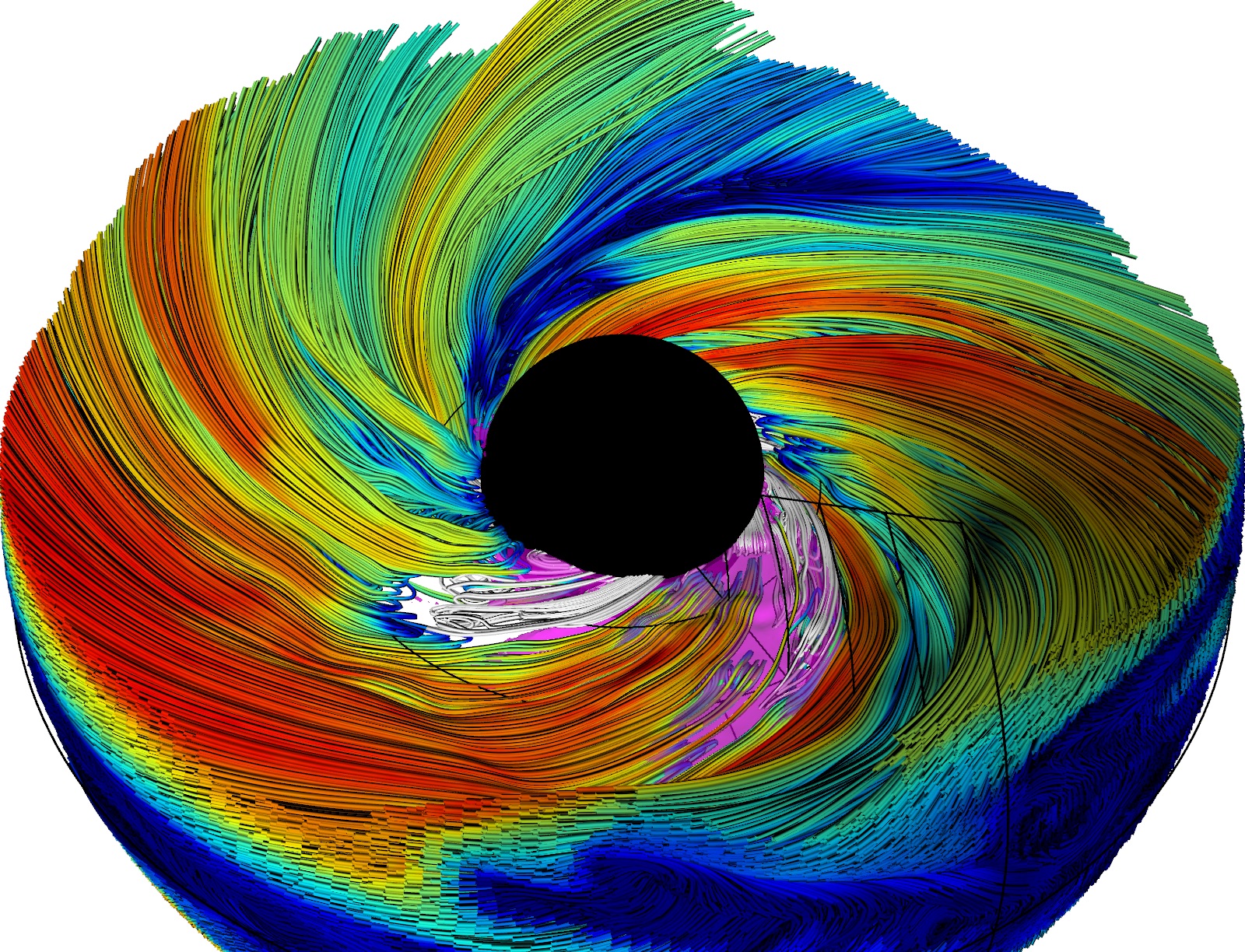}
\includegraphics[width=\columnwidth,bb=0 0 1600 1224]{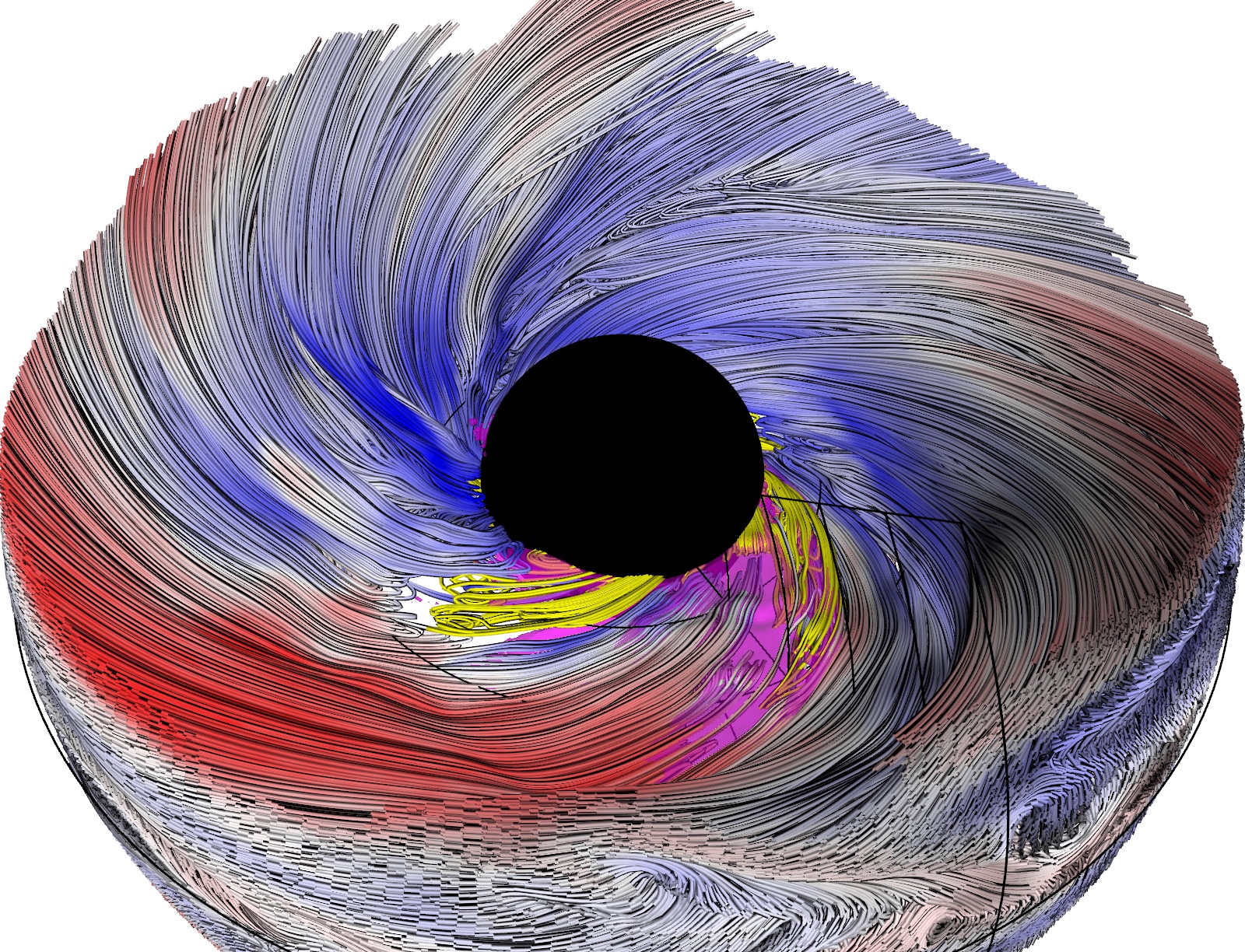}
\caption{Magnetic field lines disconnected from the BH horizon within $r \le 6 M$ at an early advanced stage of magnetic flux eruption ($t = 19630 M$) in the prograde $a = 0.9$ case.
Upper panel: lines colored by the magnetization parameter $\sigma = b^2/w$;
lower panel: lines colored by the radial velocity $v^r$.
A sample of doubly connected lines seeded at $r_0 = 1.42 M$ (just under the horizon) are shown colored in white (upper panel) or yellow (lower panel).
The magenta patches mark regions of relativistic temperature ($\log_{10}T \ge 0.5$).
The view point is at $\theta_{\rm obs} = 45^\circ$, $\phi_{\rm obs} = 180^\circ$.
Black sphere indicates the outer horizon at $r_{\rm H} = 1.436 M$.
The grid of black arcs marks the radii $r/M = 2,3,4,5,6$ in the equatorial plane $\theta = 90^\circ$, and in the $\phi = 210^\circ$ quadrant spanning the latitudes $\theta_{\rm L3,min} \le \theta \le \theta_{\rm L3,max}$.
Associated movie is available online.}
\label{fig_lines_a09_advan}
\end{figure}

Figure \ref{fig_lines_a09_advan} shows disconnected magnetic field lines for the prograde $a = 0.9$ case at $t = 19630 M$, in an early advanced stage of that magnetic flux eruption, colored by the magnetization $\sigma$ (upper panel) or by the radial velocity $v^r$ (lower panel).
An extended region of relativistic temperature can be seen around the front side of the BH horizon (while other sectors of the horizon are accreting regularly), strongly flattened along the equatorial plane.
Considering that this scene is shown from the azimuth of $\phi_{\rm obs} = 180^\circ$, this is consistent with the double-tongue-shaped region of relativistic temperature shown in the lower middle panel of Figure \ref{fig_rphmaps}.
Parts of that relativistically hot region are filled with bunches of doubly connected field lines, other parts are filled with disconnected field lines, the remaining exposed gaps strongly suggest that anti-parallel field lines from the upper and lower sides of the split-monopole horizon-connected magnetosphere come into direct contact (see the lower panels of Figure \ref{fig_rthmaps_rmax6}).
This results in relativistic reconnection with very weak guide field, which is optimal for efficient conversion of magnetic energy into heat, motion and non-thermal acceleration of particles \citep[e.g.,][]{2017ApJ...843L..27W}.
The magnetic field lines crossing the region of relativistic temperature develop fast radial outflows that reach the $r = 6 M$ radius over a broad range of $\phi$ values.
We also find much higher (as compared with Figure \ref{fig_lines_a09_onset_sigma}) magnetization values for horizon-disconnected field lines, not only for those connected to the main eruption region, but mainly for those most elevated from the equatorial plane.
Those lines may have been depleted of mass while horizon-disconnected due to velocity divergence associated with stagnation, alternatively they had been depleted while horizon-connected and have just disconnected.

\subsubsection{Magnetic field lines: short-term dynamics}
\label{sec_res_pro_lines_seq}

\begin{figure*}
\centering
\begin{tabular}{m{0.18\textwidth}m{0.18\textwidth}m{0.18\textwidth}m{0.18\textwidth}m{0.18\textwidth}}
\centering $t = 19592 M$ &
\centering $t = 19593 M$ &
\centering $t = 19594 M$ &
\centering $t = 19595 M$ &
\centering $t = 19596 M$
\end{tabular}
\includegraphics[width=\textwidth,bb=0 0 1500 300]{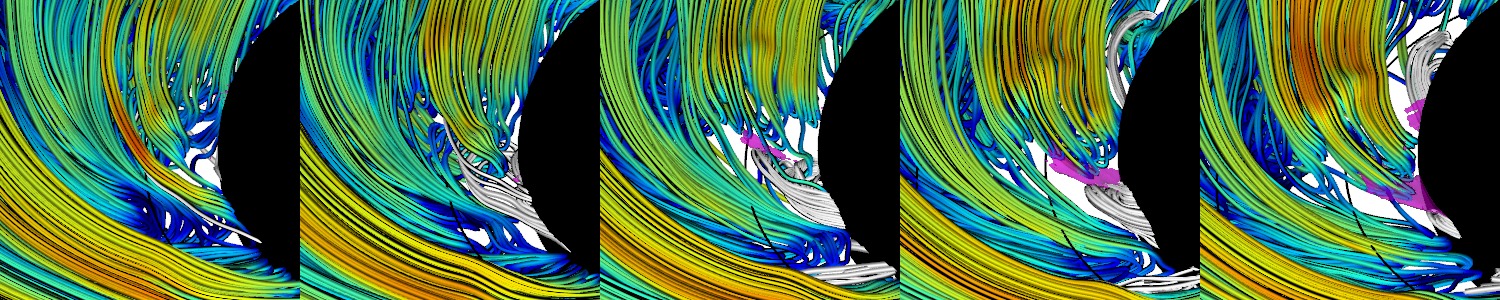}
\\
\begin{tabular}{m{0.18\textwidth}m{0.18\textwidth}m{0.18\textwidth}m{0.18\textwidth}m{0.18\textwidth}}
\centering $t = 19602 M$ &
\centering $t = 19609 M$ &
\centering $t = 19616 M$ &
\centering $t = 19623 M$ &
\centering $t = 19630 M$
\end{tabular}
\\
\includegraphics[width=\textwidth,bb=0 0 2250 450]{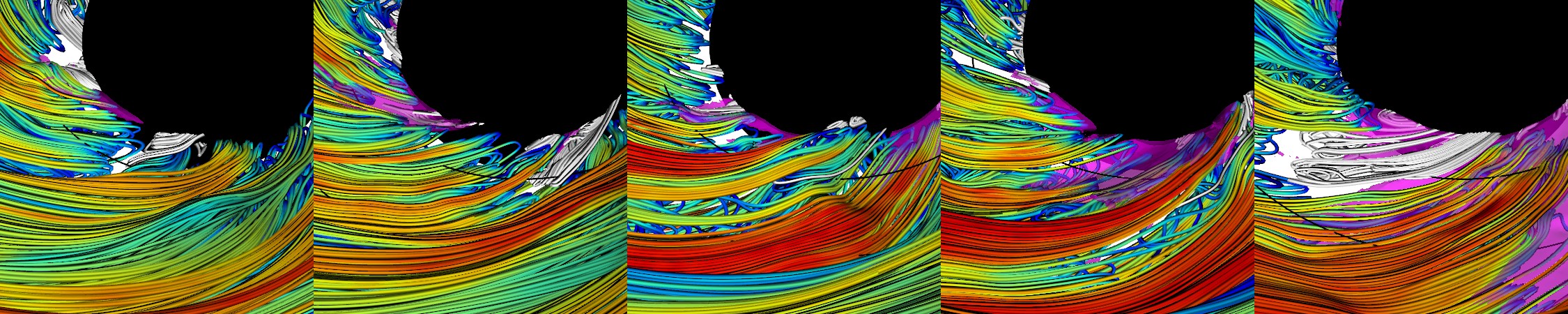}
\caption{Time evolution of magnetic field lines during the onset of magnetic flux eruption in the prograde $a = 0.9$ case.
Horizon-disconnected lines are colored by the magnetization parameter $\sigma = b^2/w$,
and doubly connected lines are shown colored in white.
The magenta patches mark relativistic temperature ($\log_{10}T \ge 0.5$).
All images are seen from the latitude of $\theta_{\rm obs} = 45^\circ$ and the azimuth of $\phi_{\rm obs} = 180^\circ$.
Black spheres indicate the outer horizon at $r_{\rm H} = 1.436 M$, and black arcs seen in most panels mark the radius $r = 2 M$ in the equatorial plane $\theta = 90^\circ$.
The upper sequence covers the time period of $t = 19592M - 19596M$ every $\Delta t = 1M$ (the middle frame corresponds to Figure \ref{fig_lines_a09_onset_sigma}).
The lower sequence covers the time period of $t = 19602M - 19630M$ every $\Delta t = 7M$ (the last frame corresponds to Figure \ref{fig_lines_a09_advan}).
Associated movie is available online.}
\label{fig_lines_seq}
\end{figure*}

Figure \ref{fig_lines_seq} presents two short-term time sequences of zoomed and cropped plots similar to Figures \ref{fig_lines_a09_onset_sigma} and \ref{fig_lines_a09_advan}, with disconnected magnetic field lines colored by the magnetization $\sigma$.

The first sequence covers the period of $t = 19592 M - 19596 M$ with time resolution $\Delta t = 1 M$.
The third (middle) frame for $t = 19594 M$ corresponds to Figure \ref{fig_lines_a09_onset_sigma} and is identified as the very onset of this magnetic flux eruption, for the reason that this is the first occurrence of a consistent patch of relativistic temperature (cf. the upper right panel of Figure \ref{fig_tphmaps}).
This hot patch appears between a small bundle of doubly connected field lines and a broad band of relatively highly magnetized disconnected lines.
The second and third frames ($t = 19593 M, 19594 M$) show those groups of field lines partially interlocked.
The doubly connected bundle appears to emerge spontaneously between the first ($t = 19592 M$) and second ($t = 19593 M$) snapshots.
By the fifth (last) frame for $t = 19596 M$, the doubly connected field lines collapse into the BH, and the high-temperature patch appears attached to the horizon.

The second sequence covers the period of $t = 19602 M - 19630 M$ with the time resolution of $\Delta t = 7 M$, showing further development of this magnetic flux eruption.
The fifth (last) frame for $t = 19630 M$ corresponds to Figure \ref{fig_lines_a09_advan}.
The first two frames ($t = 19602 M, 19609 M$) show additional bundles of doubly connected field lines appearing at different azimuths along the equatorial part of the horizon followed by more extended high-temperature region in the fourth frame ($t = 19623 M$). Isolated bands of very highly magnetized disconnected field lines appear in the third and fourth frames ($t = 19616 M, 19623 M$).

\subsection{Magnetic field lines: retrograde case}
\label{sec_res_retro}

\begin{figure}
\includegraphics[width=\columnwidth,bb=0 0 1224 1224]{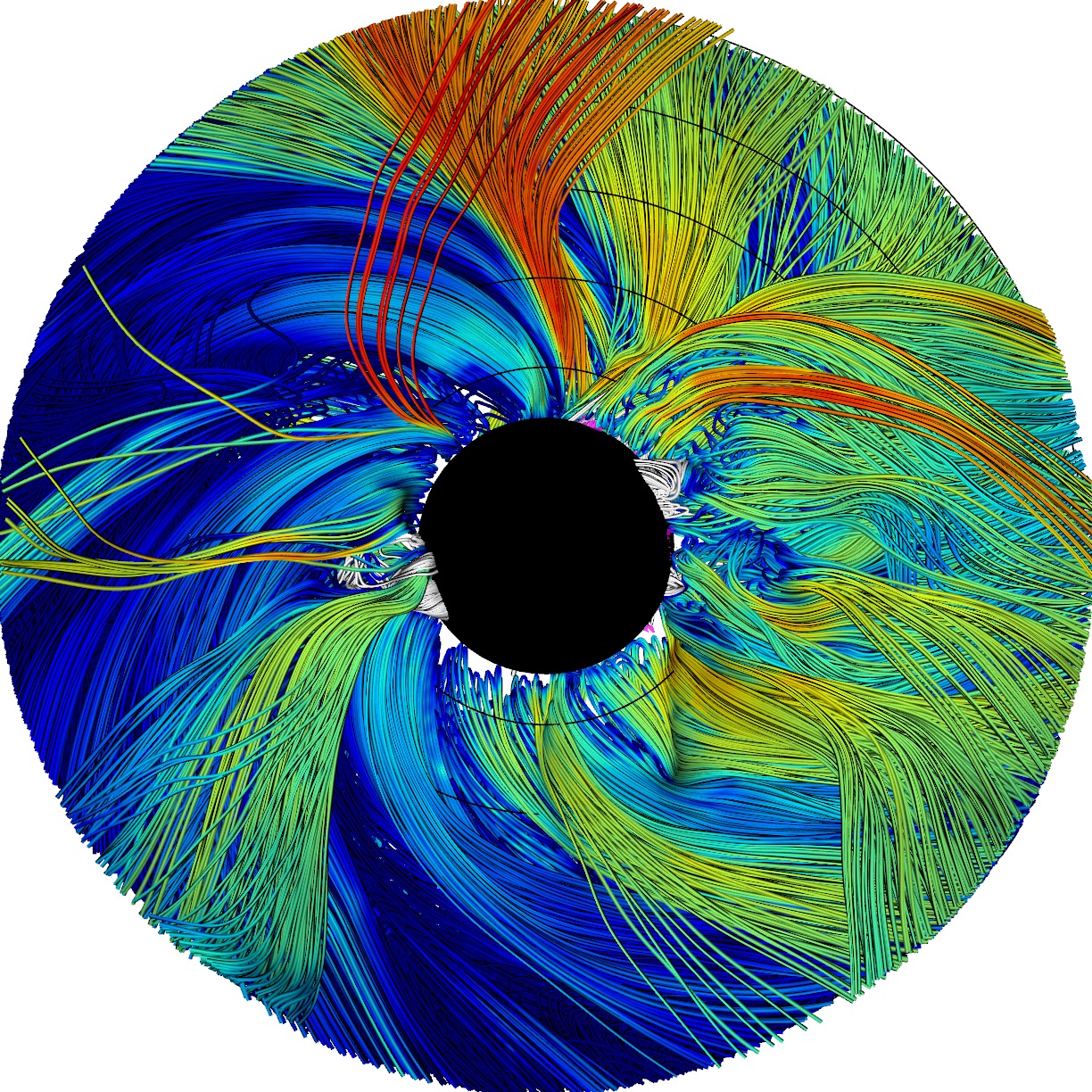}
\caption{Magnetic field lines disconnected from the BH horizon within $r < 6 M$ in the retrograde $a = -0.9$ case at the peak of BH magnetic flux during the onset of magnetic flux eruption ($t = 17050 M$; cf. Figure \ref{fig_hist_short}), with lines colored by the magnetization parameter $\sigma = b^2/w$.
A sample of doubly connected lines is shown colored in white.
The small magenta patches mark regions of relativistic temperature ($\log_{10}T \ge 0.5$).
The view point is at $\theta_{\rm obs} = 0^\circ$ with $\phi$ measured counterclockwise from the right.
Black sphere indicates the outer horizon at $r_{\rm H} = 1.436 M$.
The grid of black arcs marks the radii $r/M = 2,3,4,5,6$ in the equatorial plane $\theta = 90^\circ$.
Associated movie is available online.}
\label{fig_lines_a-09}
\end{figure}

Figure \ref{fig_lines_a-09} presents a complete sample of disconnected magnetic field lines within $r \le 6M$ colored by the magnetization $\sigma$ in the retrograde $a = -0.9$ case at $t = 17050 M$ (the peak of $\Phi_{\rm BH}$ indicated in Figure \ref{fig_hist_short}).
We also show a complete sample of doubly connected field lines.
Those lines were seeded in the same way as in the prograde case presented in Figure \ref{fig_lines_a09_onset_sigma}.

The structure of disconnected field lines in the retrograde case is more complex than in the prograde case.
The main difference is that the accretion flow has opposite sense of angular momentum (in both cases, the BH spin vector points towards $\theta = 0$; to the observer in Figure \ref{fig_lines_a-09}).
In the retrograde case, the angular momentum of the initial torus, and consequently of the magnetically disconnected accretion flow, points towards $\theta = 180^\circ$ (away from the observer in Figure \ref{fig_lines_a-09}).

Low-$\sigma$ (blue) lines in Figure \ref{fig_lines_a-09} trace a regular accretion flow with azimuthal twist such that $B^r B^\phi > 0$, opposite to the prograde case.
The upper side of Figure \ref{fig_lines_a-09} ($\phi \sim 90^\circ$) shows a bundle of high-$\sigma$ (orange) lines forming a fan structure indicating a smoothly reversing azimuthal twist.
In radial velocity, those reversing lines show stagnation or weak outflow ($v^r \gtrsim 0$).
A minor fan structure can also be seen rooted at $\phi \sim 120^\circ$, with sparse high-$\sigma$ (red) lines extending to the foreground.
Most of the connected field lines show reversed azimuthal twist with $B^r B^\phi < 0$, like in the prograde case.
Another band of high-$\sigma$ lines can be seen rooted over the range of $\phi \sim 60^\circ - 90^\circ$ and swept to the right.

The right half of Figure \ref{fig_lines_a-09} ($-90^\circ < \phi < 90^\circ$) shows more green and yellow color, which indicates $\log_{10}\sigma \sim 0 - 0.5$.
A minor patch or relativistic temperature seen just outside the horizon at $\phi \simeq 80^\circ$ is the actual seed of magnetic flux eruption that is going to develop intermittently (cf. the animated version of Figure \ref{fig_lines_a-09}) and eventually engulf more than half of the horizon circumference (cf. the lower right panel of Figure \ref{fig_tphmaps}).
It is not such a clean eruption trigger event as the one in the prograde case illustrated in Figure \ref{fig_lines_seq}, since it was preceded by intermittent activity over at least $\Delta t \sim 170 M$, during which the global $\Phi_{\rm BH}$ remained roughly stable.

On the opposite side of the BH, at $\phi \sim 210^\circ$, a thick bundle of doubly connected lines wrap around a comparably thick tube of medium-$\sigma$ (green) disconnected lines.
The animated version of Figure \ref{fig_lines_a-09} shows that this is a final stage of rapid ($\Delta t \sim 5 M$) advection of a large concentration of magnetic flux onto the BH, which is responsible for a sharp increase of $\Phi_{\rm BH}$ shown in Figure \ref{fig_hist_short} (directly preceding the blue circle).

In the lower right quadrant of Figure \ref{fig_lines_a-09}, at $r \simeq 2 M$ and $\phi \simeq -45^\circ$, one can notice a warp in the distribution of disconnected lines, causing them to bend obliquely not only in the equatorial plane, but also across.
This pattern propagating retrograde with the accretion flow (cf. the animated version of Figure \ref{fig_lines_a-09}) is likely an oblique shock (or at least a contact discontinuity).
In the foreground of its left side (depressed under the equatorial plane) is located a coherent tube of connected field lines (not shown here) oriented parallel to the discontinuity.
Such structures appear to be more common in our retrograde simulations than in the prograde cases (including lower resolutions).

\subsection{$(t,\phi)$ maps}
\label{sec_res_tphmaps}

\begin{figure*}
\begin{center}
\includegraphics[width=\graphreduce\textwidth,bb=0 0 720 288]{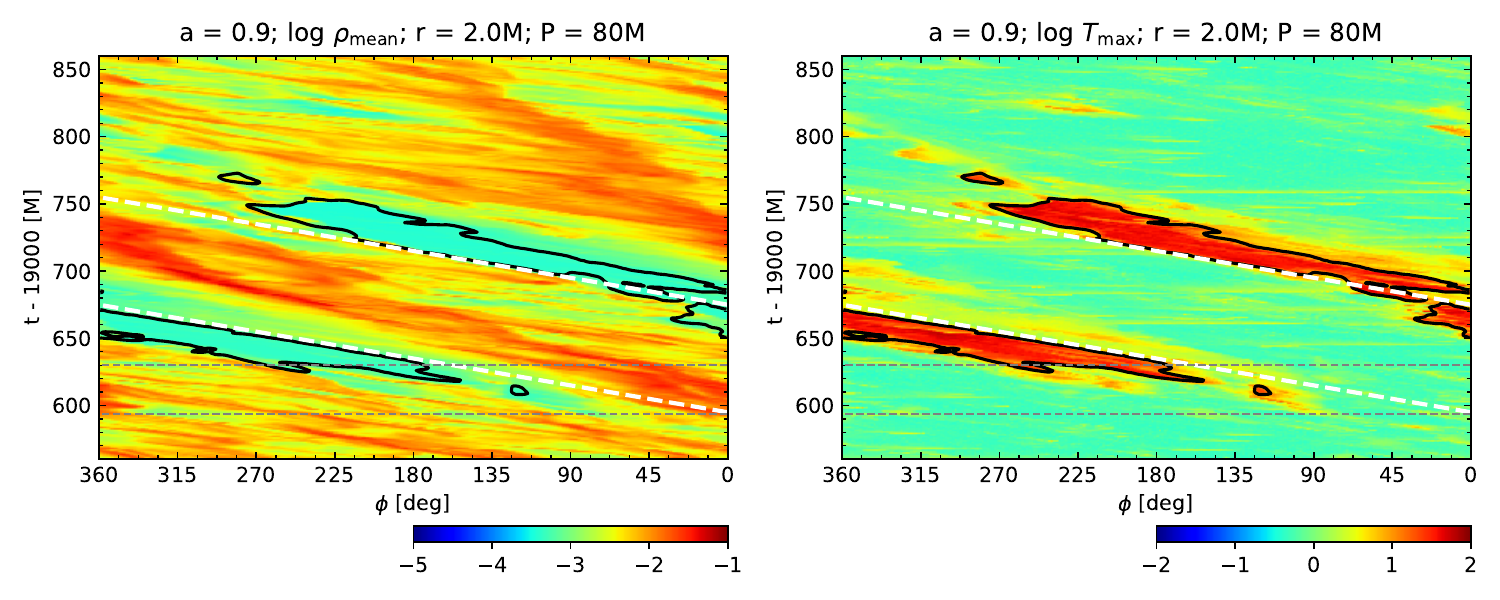}
\includegraphics[width=\graphreduce\textwidth,bb=0 0 720 288]{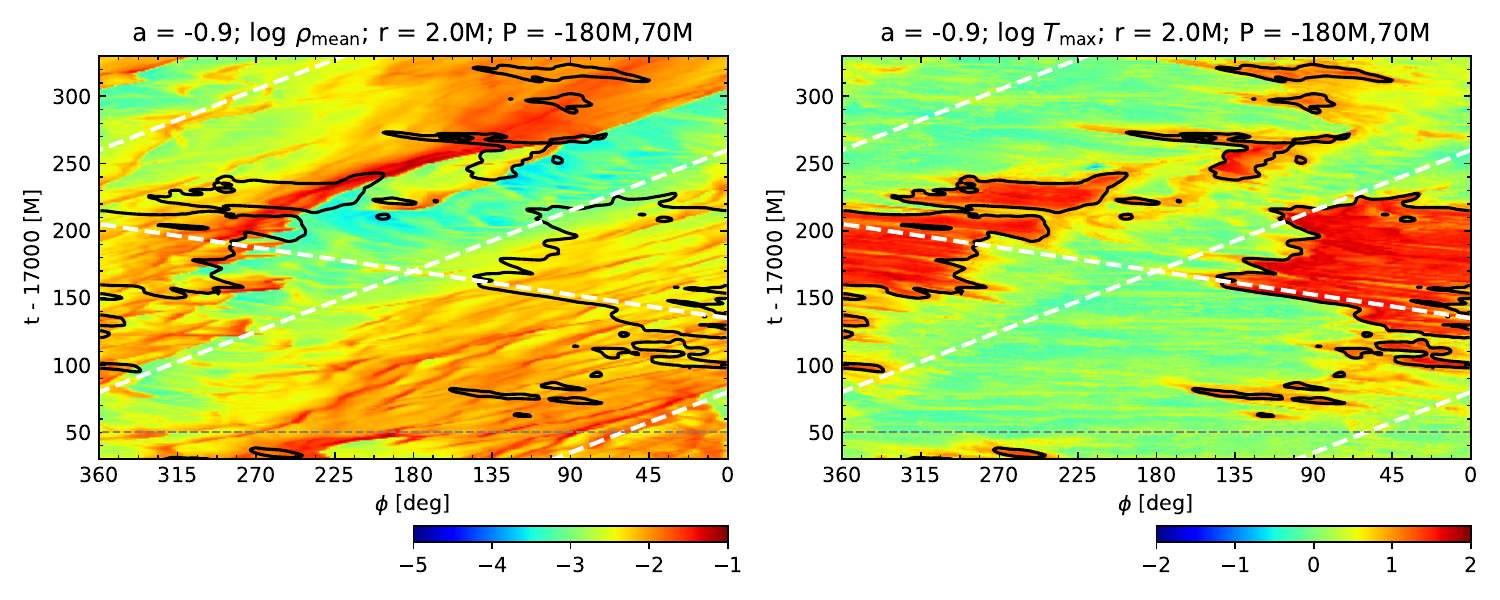}
\end{center}
\caption{Spacetime diagrams in the $(t,\phi)$ coordinates for parameters presented in Figure \ref{fig_rphmaps}: mean plasma density $\rho_{\rm mean}$ (left panels) and maximum plasma temperature $T_{\rm max}$ (right panels), with statistics taken over $\theta_{\rm L3,min} < \theta < \theta_{\rm L3,max}$ at fixed radius $r = 2 M$.
The black contours indicate $\log_{10}T_{\rm max} = 1$.
The white dashed lines mark linear rotation trends.
The upper panels present the prograde $a = 0.9$ case in the time window $19500 M < t < 19900 M$ (trend with rotation period $85 M$);
the lower panels present the retrograde $a = -0.9$ case in the time window $16900 M < t < 17400 M$ (trends with rotation periods $-180 M, 70 M$).
}
\label{fig_tphmaps}
\end{figure*}

Figure \ref{fig_tphmaps} shows spacetime diagrams of $\theta$-averaged plasma density $\rho_{\rm mean}$ and $\theta$-maximized plasma temperature $T_{\rm max}$ extracted at radius $r = 2 M$ and presented as functions of coordinates $(t,\phi)$.
This is based on the $(r,\phi)$ maps similar to those presented in Figure \ref{fig_rphmaps}, but for many additional time steps.
We compare magnetic flux eruptions in the prograde $a = 0.9$ and retrograde $a = -0.9$ cases.

The upper panels of Figure \ref{fig_tphmaps} present an eruption in the prograde case in the time window $19560 M \le t \le 19860 M$.
We indicate two particular epochs (Figure \ref{fig_hist_short}): the eruption onset at $t = 19594 M$ and $\phi \simeq 90^\circ$ (presented in Figures \ref{fig_rthmaps_rmax18}, \ref{fig_rthmaps_rmax6} - upper panels, \ref{fig_rphmaps} - upper panels, and particularly \ref{fig_lines_a09_onset_sigma} and \ref{fig_lines_a09_onset_vr}), and an early advanced eruption stage at $t = 19630 M$ and $180^\circ \lesssim \phi \lesssim 270^\circ$ (presented in Figures \ref{fig_rthmaps_rmax6} - lower panels, \ref{fig_rphmaps} - lower panels, and particularly \ref{fig_lines_a09_advan}).
The region of relativistic temperature coincides closely with the density gap.
The main region of relativistic temperature exists for $19620 M < t < 19760 M$ and performs a coherent rotation by $\Delta\phi = +480^\circ$.
We present a trend line of linear rotation with the period of $P = +85 M$, which also indicates a glitch at $\phi \simeq 30^\circ$ for $t \simeq 19680 M$.
Besides the eruption, the density map shows a large number of short-lived high-density filaments performing coherent rotations at similar rates.

The lower panels of Figure \ref{fig_tphmaps} present an eruption in the retrograde case in the time window $17030 M \le t \le 17330 M$.
We indicate one particular epoch (Figure \ref{fig_hist_short}) at $t = 17050 M$, presented in Figure \ref{fig_lines_a-09}.
In this case, the regions of relativistic temperature do not coincide with the density gap (they actually appear to avoid it).
The main density gap performs a coherent rotation by $\Delta\phi \simeq -525^\circ$ over $17070 M < t < 17370 M$.
At the epoch of $t = 17050 M$, it appears to be limited to $270^\circ \lesssim \phi \lesssim 360^\circ$, the lower right quadrant of Figure \ref{fig_lines_a-09} (including the warp at $\phi \simeq 320^\circ$).
We plot a trend line of linear rotation towards decreasing $\phi$ (against the spacetime drag) with the period of $-180 M$.
This trend is similar to the rotation patterns of the high-density filaments.
The regions of relativistic temperature do not follow such a rotation trend.
The main hot region appears largely stationary in $\phi$, however, some of its edges and other short-lived features suggest a fast rotation towards increasing $\phi$ (along the spacetime drag).
We plot a second trend line for linear rotation towards increasing $\phi$ with the period of $70 M$.
We suggest the following interpretation of these results: as the density gap corotates with the retrograde accretion flow, reconnection is triggered along the trailing edge of the gap and spreads along the magnetic field lines rotating along the spacetime drag (field lines that have just disconnected from the BH horizon).

\subsection{Alignment between velocity and magnetic fields}
\label{sec_res_vperp}

We decompose velocity 3-vectors $\vec{v} = \vec{v}_\parallel + \vec{v}_\perp$ into components parallel and perpendicular to the local magnetic field 3-vectors $\vec{B}$, using the spatial metric dot product $\vec{v}\cdot\vec{B} = v_i B^i = g_{ij} v^i B^j$ and analogous vector norms $|\vec{v}| = (g_{ij} v^iv^j)^{1/2}$ and $|\vec{B}| = (g_{ij} B^iB^j)^{1/2}$.
We introduce the cosine of pitch angle $\mu(\vec{v},\vec{B}) = (\vec{v}\cdot\vec{B}) / (|\vec{v}| |\vec{B}|) \in [-1:1]$, calculate the parallel velocity $v_\parallel^i = (\vec{v}\cdot\vec{B})B^i/|\vec{B}|^2 =  \mu(\vec{v},\vec{B})\,|\vec{v}| B^i/|\vec{B}|$, and the perpendicular velocity $\vec{v}_\perp = \vec{v} - \vec{v}_\parallel$.
In ideal MHD, perpendicular velocity is a signature of electric fields.

\begin{figure*}
\begin{center}
\includegraphics[width=\graphreduce\textwidth,bb=0 0 1080 360]{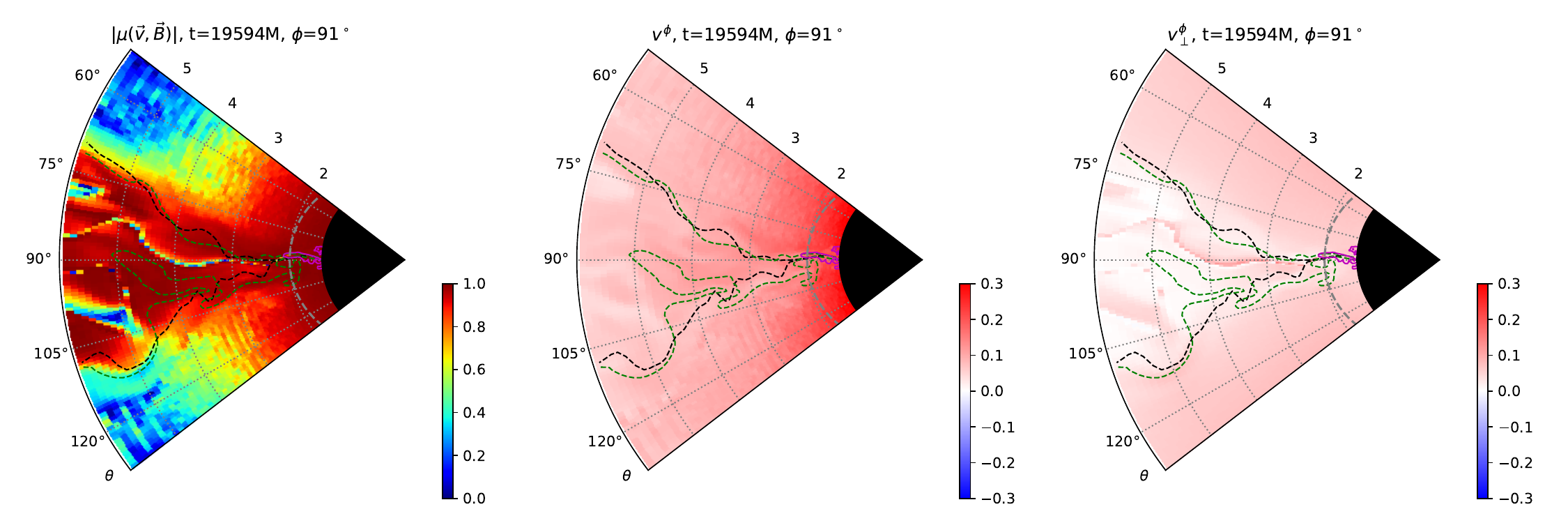}
\includegraphics[width=\graphreduce\textwidth,bb=0 0 1080 360]{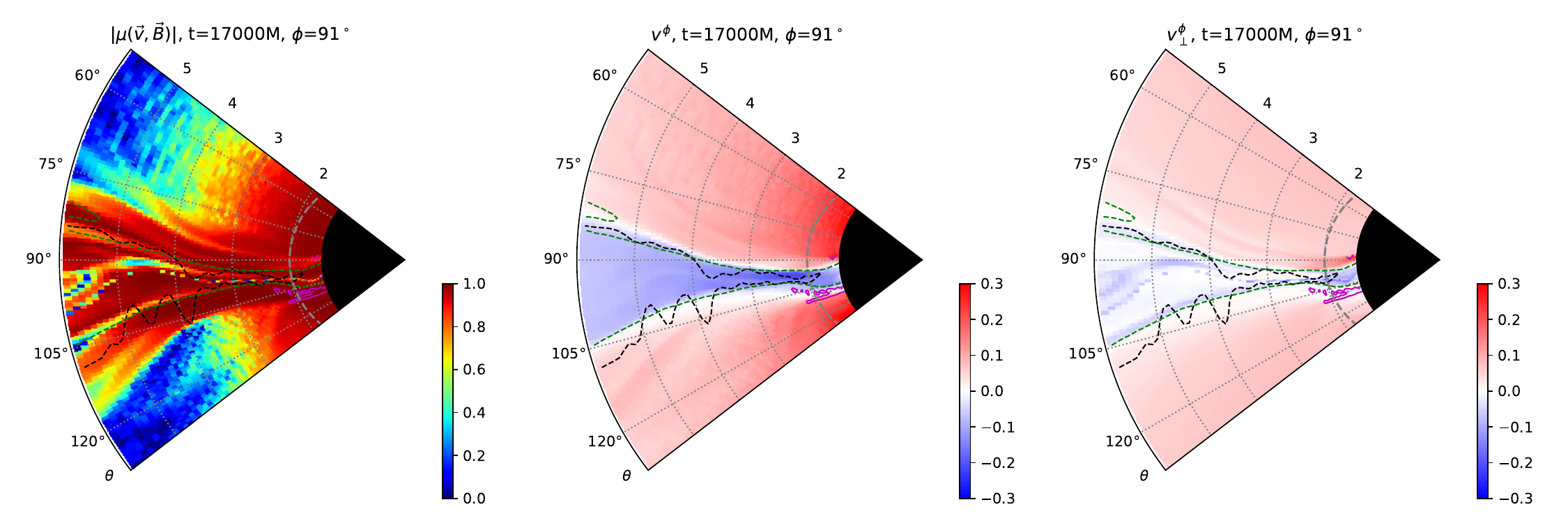}
\end{center}
\caption{Maps in the $(r,\theta)$ coordinates showing (from the left):
absolute value of the cosine of the pitch angle between the velocity 3-vector $v^i$ and the magnetic field 3-vector $B^i$;
azimuthal component of the velocity 3-vector $v^i$;
azimuthal component of the velocity 3-vector $v_\perp^i$ perpendicular to the local $B^i$.
Upper row of panels: the prograde $a = 0.9$ case for $t = 19594 M$ and $\phi \simeq 90^\circ$;
lower row of panels: the retrograde $a = -0.9$ case for $t = 17000 M$ and $\phi \simeq 90^\circ$.}
\label{fig_rthmaps_vperp}
\end{figure*}

Figure \ref{fig_rthmaps_vperp} presents in the $(r,\theta)$ coordinates the absolute value of pitch cosine $|\mu(\vec{v},\vec{B})|$, as well as the total and perpendicular azimuthal velocities $v^\phi$ and $v^\phi_\perp$, for both prograde and retrograde cases.
In the high-density accretion flow, the values of $|\mu(\vec{v},\vec{B})| \simeq 1$ indicate that the plasma velocity is closely parallel to the local magnetic field.
The exception to this are the low values of $|\mu(\vec{v},\vec{B})|$ along the equatorial current layer.
Thus, accreting plasma is generally forced to flow along the magnetic field lines, except at their sharp inner tips rooted at the current layer.
Because most disconnected field lines have a universal shape reflecting the dominance of $B^r,B^\phi$ components (see Figure \ref{fig_lines_examples}), plasma flowing radially inwards ($v^r < 0$; see Figure \ref{fig_rthmaps_rmax6}) is channeled towards the equatorial current layer, this explains the strong radial stratification of plasma density.
The inner tips of disconnected field lines might form obstacles (chokepoints) for further accretion, however, breaking of the ideal MHD flux-freezing condition, as well as stochastic variations in time (warping, flapping, etc) of the current layer due to turbulent perturbations, allow for efficient cross-field diffusion.

The azimuthal component of perpendicular velocity $v_\perp^\phi$ is positive for almost every $(r,\theta)$, and significantly smaller in the disconnected accretion flow than in the equatorial current layer and in the connected magnetosphere.
It is also significantly smaller in both regions than $v^\phi$, which within the ergosphere approaches the BH angular velocity $\Omega_{\rm H} = a/(2r_{\rm H}) = 0.313$ (for $a = 0.9$).

The disconnected field lines appear thus to be more static than the plasma flowing along them. The azimuthal component of perpendicular velocity indicates a slow rigid rotation, not sensitive to the differential rotation of space-time in the vicinity of the ergosphere.
The disconnected lines appear as stiff avenues for the accreting plasma, despite low $\sigma$ and high $\beta$, they seem to rule the motion of plasma, rather than being advected and tangled by that motion.
A major exception to this is the equatorial current layer, where ideal MHD conditions are violated and numerical cross-field diffusion prevents choking the accretion flow.

In the retrograde case, accreting plasma with negative radial velocity $v^r < 0$ flows along the disconnected magnetic field lines with $B^r B^\phi > 0$, which implies negative azimuthal velocity $v^\phi < 0$.
The perpendicular component $v_\perp^\phi$ for the accretion flow is also negative, but much closer to zero.
In the magnetosphere, where $B^r B^\phi < 0$, the azimuthal velocity is positive $v^\phi > 0$, while radial velocity transitions from an inflow ($v^r < 0$) for $r \lesssim 3 M$ (along the field lines) to an outflow ($v^r > 0$) for $r \gtrsim 3 M$ (across the field lines).

\section{Discussion}
\label{sec_disc}

\subsection{Magnetic arresting}

In the model of Magnetically Arrested Disk (MAD) \citep{2003PASJ...55L..69N}, strong latitudinal (`vertical') magnetic field $B^\theta$ forms a magnetic barrier across the equatorial plane that arrests the accretion flow.
This model postulates a radial stratification of the plasma parameter $\beta = P/P_{\rm mag}$, the ratio of plasma pressure $P$ to the magnetic pressure $P_{\rm mag} = b^2/2$, at certain magnetospheric radius $R_{\rm m}$, such that $\beta < 1$ for $r < R_{\rm m}$ and $\beta \gg 1$ for $r > R_{\rm m}$ (satisfying the balance of total pressure across $R_{\rm m}$).
The MAD model has been demonstrated in 2D (axisymmetric) simulations \citep[e.g.,][]{2008ApJ...677..317I,2021MNRAS.508.1241C}, in which accretion states accumulating magnetic flux on the BH horizon alternate with eruption states ejecting BH magnetic flux via relativistic magnetic reconnection.

In the 3D case, it has been postulated by \cite{2003PASJ...55L..69N} that accretion could proceed through the magnetic barrier in the form of filaments resulting from the magnetic Rayleigh-Taylor (interchange) instability \citep[e.g.,][]{2007ApJ...671.1726S}.
Such filaments can indeed be realized in a rigid dipolar magnetosphere rooted at a star \citep[e.g., protostar or neutron star;][]{2008MNRAS.386..673K,2024MNRAS.528.2883Z,2023arXiv231104291P} --- such magnetosphere forms a robust magnetic barrier.
In the case of BH accretion, a proper context for the interchange modes are magnetic flux tubes ejected by magnetic flux eruptions, as has been recently demonstrated by \cite{2023PhRvR...5d3023Z}.
The ejected flux tubes are magnetically dominated ($\beta < 1$) structures of strong latitudinal field that locally arrest (or repel; \citealt{2023ApJ...943L..29H}) the plasma dominated ($\beta > 1$) accretion flow, which results in the development of dense filaments within the flux tube.

3D simulations of accretion onto magnetically saturated BHs also show strong azimuthal (along $\phi$) structures in the form of `spiral arms' of plasma density \citep[e.g.,][]{2012MNRAS.423.3083M}. However, such structures are not consistent with interchange modes\footnote{Geometrically thin accretion flows can be warped, so that a low-density magnetosphere may imprint an azimuthal structure in the equatorial plane that could be attributed to the interchange instability (this appears to be the case in Figure 4 of \citealt{2012MNRAS.423.3083M}). For this reason our Figure \ref{fig_rphmaps} presents $(r,\phi)$ maps of parameter statistics drawn from a range of $\theta$ values.}: (1) the range of plasma parameter values is typically limited to $\beta > 1$; (2) radial velocity is roughly uniform across the density arms.

Most GRMHD simulations of magnetized accretion onto BHs (at least since the works of \citealt{2003ApJ...589..444G,2003ApJ...589..458D}) have been initiated from geometrically thick toroidal hydrodynamical equilibria \citep{1976ApJ...207..962F,1985ApJ...288....1C} seeded with weak magnetic fields (most often poloidal; but recently also toroidal, \citealt{2020MNRAS.494.3656L}), such that initial plasma $\beta \gtrsim 100$ and accretion results from the magnetorotational instability (MRI; \citealt{1991ApJ...376..214B}).
The inner parts of the poloidal field loops are advected onto the BHs by largely stable and axisymmetric accretion flows maintaining high values of plasma $\beta$, i.e. dominated by plasma pressure.
At no point there is an opportunity to develop a radial stratification of $\beta$ and hence a synchronized magnetic barrier.
The magnetic flux through the equatorial plane is much smaller than the magnetic flux collected across the BH horizon (\citealt{2022MNRAS.511.2040B}).
Also in our simulations we find that the latitudinal (vertical) magnetic field component $B^\theta$ is typically one order of magnitude weaker than the radial and toroidal components $B^r,B^\phi$, even when the BH is magnetically saturated and the inner accretion flow is geometrically thin.

Our simulations do not show any evidence for thin accretion flows breaking into azimuthal filaments.
Such accretion flows pass smoothly all the way to the BH horizon, avoiding obstacles that might require developing instabilities like interchange to be overcome.
A potential chokepoint upon approaching the equatorial current layer, where accreting plasma reaches the inner tip of its magnetic field line, can be avoided due to cross-field diffusion, which allows accretion to continue along the current layer.

The key to achieving a true MAD configuration in 3D appears to be synchronization of the magnetic barrier.
Early simulations presented by \cite{2008ApJ...677..317I} were initiated in 2D and continued only briefly in 3D, with a magnetic barrier established in the 2D stage.
In longer 3D simulations of geometrically thick accretion flows (\citealt{2011MNRAS.418L..79T,2012MNRAS.423.3083M,2022ApJ...924L..32R}; and many other studies, including this work),
occasionally ejected magnetic flux tubes form temporary and localized magnetic barriers, which can hardly be extended synchronously over all $\phi$ values.
A true MAD appears to have been achieved in the radiative GRMHD simulations with poloidal magnetic fields described in the work of \cite{Liska22}, with radiative cooling resulting in vertical collapse of the initial geometrically thick torus into a geometrically thin Keplerian disk with the inner edge truncated at $r \simeq 25 M$ by a proper barrier of `vertical' magnetic fields.
This is not inconsistent with our interpretation of geometrically thick RIAFs.

\subsection{Magnetic choking}

The bulk of accretion flows onto magnetically saturated BHs do not appear to be magnetically choked in the sense proposed by \cite{2012MNRAS.423.3083M}.
`Choked' suggests that it is difficult for the accretion flow to proceed into a converging nozzle between two magnetospheres.
This term can be applied only to the part of accretion flow directly upstream from a forming magnetic X-point.
However, radial profiles of plasma density, averaged over $t,\theta,\phi$, in accretion flows onto magnetically saturated BHs, are consistent with a power law $\rho \propto r^{-1}$ (\citealt{2022MNRAS.511.2040B}), with no evidence for a break that would indicate an additional compression.
Latitudinal stratification of plasma density appears to be regulated by the disconnected magnetic field lines.
The disconnected lines are in direct contact and in latitudinal pressure balance with the connected lines of the magnetosphere, and yet they remain obliquely elevated from the densest layer of accreting plasma.
There is a smooth continuity between the disconnected and connected lines, which together form a twisted split monopole.
Hence, even in a saturated state the BH magnetosphere does not directly compress the accreting plasma.

Moreover, such accretion flows do not appear to be affected by an angular momentum barrier, since even in such narrow nozzles there is a room for laminar winds that provide enough magnetic torque to maintain these flows in a sub-Keplerian plunge (\citealt{2024MNRAS.527.1424S}; see also \citealt{2024ApJ...965..175M}).
Magnetic flux eruptions also contribute to the outwards angular momentum transport \citep{2018MNRAS.478.1837M}, even for non-spinning BHs \citep{2022ApJ...941...30C}.
Sub-Keplerian rotation rates characterize the accreting plasma, and even more so the disconnected magnetic field lines (Section \ref{sec_res_vperp}), and are reflected in the azimuthal patterns ray traced into BH crescent images \citep{2023ApJ...951...46C}.

\subsection{Magnetic channeling}

We find that in the narrow nozzle formed by two bulging connected magnetospheres, geometric constriction of MRI turbulence forces the disconnected magnetic field lines to organize themselves into a rigid slowly rotating structure.
Once a saturated BH cannot accept additional flux, the pressure of connected field lines prevents further advection of disconnected lines.
Additional analysis of trajectories of test particles (Appendix \ref{app_test_part}) indicates that
the disconnected lines channel the accretion flow towards the equatorial current layer, resulting in the latitudinal density stratification.
The accretion flow does not stop at the current layer, but proceeds along it by turbulent cross-field diffusion.
The equatorial current layer thus forms the ultimate channel for the accretion flow to reach the BH.
Such accretion flows are thus effectively magnetically channeled into the BH (cf. magnetic channeling into a jet/wind, \citealt{1993A&A...276..637F}).

\subsection{Regularity of disconnected field lines}

\begin{figure}
\includegraphics[width=\columnwidth,bb=0 0 1443 763]{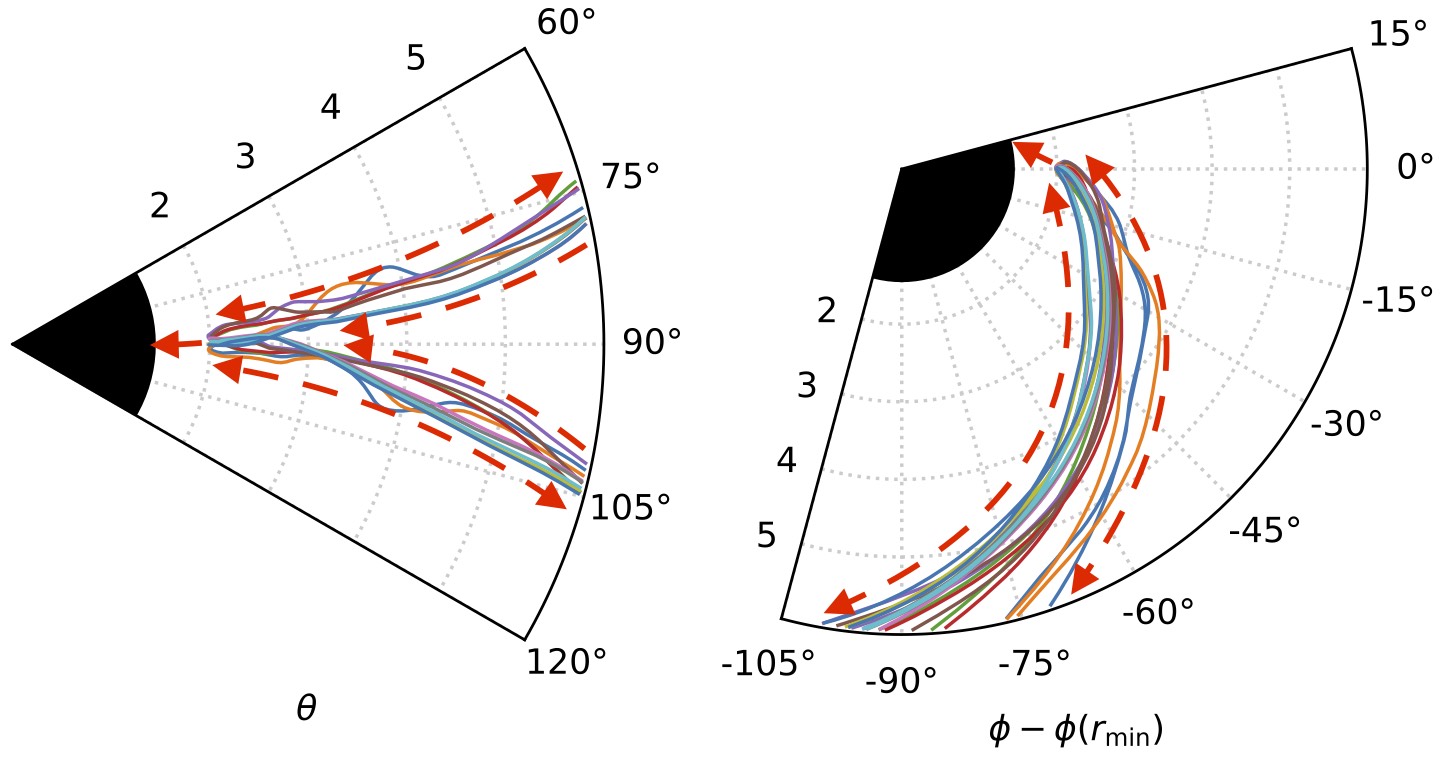}
\caption{Examples of regular field lines disconnected from the BH horizon, with inner tips at $r_{\rm min} \simeq 2 M$ aligned in $\phi$.  Red dashed arrows illustrate the channeling effect of field lines on plasma motions (accretion, and outflows for the most $\theta$-elevated lines).}
\label{fig_lines_examples}
\end{figure}

We find that the inner accretion flow is threaded by disconnected field lines attaining a universal shape illustrated in Figure \ref{fig_lines_examples} --- a double spiral with a sharp inner tip close to the equatorial plane, from which extend outwards two branches roughly symmetric with respect to the equatorial plane, systematically diverging away from that plane, and spiraling in the same azimuthal direction, reflecting the dominance of radial and toroidal components $B^r,B^\phi$.
A sharp tip means a sharp field gradient, hence a strong electric current density.
Most disconnected field lines are rooted along the single equatorial current layer.
Accreting plasma is tied to those lines, forced to move parallel to them.
The slowly rotating lines guide the plasma towards the current layer.
This is consistent with the strong latitudinal density stratification, which is realized in our simulations instead of the radial stratification postulated by the MAD model.
Should ideal MHD be satisfied strictly, the line tips would form chokepoints for the plasma.
However, if sufficient plasma density accumulates at those line tips, it would pull the line tip inwards.
On the other hand, a finite numerical resolution limits the field gradient, i.e., the current density, and the sharpness of the line tip.
Under finite resolution, ideal MHD must break down along the current layer, allowing the accumulated plasma to diffuse inwards, across the line tip (hence $v_\perp^r < 0$ along the current layer; Figure \ref{fig_rthmaps_vperp}).
In addition, the equatorial current layer is not steady, but flapping and warping due to perturbations advected from the turbulent outer zones.

This picture is consistent with the pattern of linear polarization observed in M87* by the Event Horizon Telescope \citep{2021ApJ...910L..12E}, which has been interpreted by \cite{2021ApJ...921L..38P} in terms of a Parker spiral.

\subsection{Triggering magnetic flux eruptions}

We identified the trigger event for magnetic flux eruption as activation of pre-existing equatorial current layer due to decrease of plasma density.
A necessary but insufficient condition is the change of magnetic field topology (formation of magnetic X-points).
This requires disconnection of a small group of field lines from the BH horizon.
Spontaneous appearance of doubly connected field loops suggests that disconnection is chaotic \citep[in the spirit of][]{1999ApJ...517..700L}, resulting from perturbations of the innermost geometrically constricted accretion flow, advected from larger radii where accretion is geometrically thick and MRI turbulence is fully developed.
Disconnection of magnetospheric field lines introduces low-density plasma to the disconnected domain, from which it can be transmitted to an X-point.

This scenario is qualitatively similar to the scenarios proposed to explain analogous magnetic eruptions in other astrophysical environments.
In the context of solar flares, it was demonstrated using ideal MHD simulations that gradual thinning of a current layer can non-linearly accelerate reconnection \citep{2021NatAs...5.1126J}.
In the context of Earth's magnetotail region, in situ measurements by the Magnetospheric MultiScale (MMS) constellation of probes allowed to attribute the onset of magnetic reconnection to the thinning of a current layer due to a transient thermal pressure pulse in the solar wind, rather than due to loading of additional magnetic flux \citep{2023JGRA..12831758G}.

\subsection{Self-organized criticality}

We have demonstrated in Section \ref{sec_res_pro} development of a full-scale magnetic flux eruption from a very compact reconnection site related to a brief occurrence of a doubly connected field loop.
We have considered other examples of flux eruptions, large and small --- they all start from similar conditions, essentially from scratch.
At even lower resolutions, the initial phase of magnetic flux eruption can be characterized as a self-similar growth of the relativistically hot density gap due to a positive feedback cycle of (decreasing density $\to$ increasing temperature $\to$ radial acceleration $\to$ decreasing density) \citep{2024hepr.conf}.
Most doubly connected field loops do not trigger an eruption, and there are clearly more small (failed) eruptions than large ones (Figure \ref{fig_tphmaps}).
Such conditions are generally analogous to the solar and stellar flares \citep[e.g.,][]{1991ApJ...380L..89L,2021ApJ...910...41A,2022ApJ...925L...9F}, and it is reasonable to expect that BH flux eruptions are example of the principle of self-organized criticality (a magnetic avalanche), in which small perturbations trigger events of unpredictable magnitude \citep{1988PhRvA..38..364B}.
This would predict that the magnetic energy released during BH flux eruptions should follow a power-law distribution $N(\mathcal{E}) \propto \mathcal{E}^{-p}$ with the index $p \simeq 1.5$.

\subsection{Rotation of flux eruption footpoints}

Previous studies \citep{2020MNRAS.497.4999D,2021MNRAS.502.2023P} showed that magnetic flux eruptions can eject compact tubes of vertical magnetic field orbiting the BH at radii $r \sim 10 M$, which have been applied to the flares of Sgr~A* resolved by the GRAVITY \citep{2018A&A...618L..10G}.
In Section \ref{sec_res_tphmaps} we showed that systematic rotation can be performed also by the eruption footpoint measured at $r \simeq 2 M$, and similar rotations can be demonstrated at radii $r \sim 4M$ contributing most directly to a BH crescent image.
However, low plasma density may cause the eruption footpoints to appear dark in crescent images.
Even in such a case, observation of a coherently rotating dark band in the crescent image of M87*, correlated with other multiwavelength signatures, e.g., a TeV gamma-ray flare, or a helically propagating radio-VLBI feature, would be extremely useful to constrain the theoretical picture of BH flux eruptions.

\subsection{Prograde vs. retrograde accretion flows}

Prograde and retrograde cases for magnetically saturated BH accretion flows have been compared in several studies based on GRMHD simulations \citep{2012MNRAS.423L..55T,2022MNRAS.511.3795N,2024ApJ...962..135Z}.
Although various quantitative differences have been reported, the general picture is that such cases would be difficult to distinguish observationally.
This was also the case in the interpretation of the BH crescent images obtained by the Event Horizon Telescope by comparing them with images produced from GRMHD simulations: in the case of M87*, both prograde and retrograde MAD models could satisfy the considered constraints \citep{2019ApJ...875L...5E}, including linear polarization \citep{2021ApJ...910L..13E} and circular polarization \citep{2023ApJ...957L..20E}.

Our results presented in Sections \ref{sec_res_retro} and \ref{sec_res_tphmaps} suggest that the retrograde case is much more complex, including chaotic transfer of magnetic field lines across a sharp shear layer.
This should result in much more vigorous particle acceleration than in the prograde case, and potentially in distinct observational signatures.
Global kinetic simulations using the general relativistic particle-in-cell (GRPIC) algorithm are necessary to verify this.

\section{Conclusions}
\label{sec_conc}

Using the results of 3D ideal GRMHD numerical simulations, we analyzed magnetic connectivity of magnetically saturated accreting Kerr BHs (also known as MAD or MCAF), focusing on the magnetic field lines disconnected from the BH horizon (disconnected lines), occupying the magnetically disconnected volume domain (disconnected domain), which separates two connected domains (bipolar split-monopole BH magnetosphere).

The inner part of the disconnected domain is geometrically constricted into a narrow nozzle, eventually restricting the outer magnetorotational turbulence to a single equatorial current layer, subject to significant fluctuations.
Azimuthal sectors of the disconnected domain alternate between the states of magnetic flux accumulation (regular accretion flow with low magnetization $\sigma \ll 1$ with strong latitudinal stratification and inactive equatorial current layer) and magnetic flux eruption (active equatorial current layer drives relativistic magnetic reconnection evidenced by relativistic temperature).

The surface regions of the disconnected domain are characterized by laminar magnetic fields transitioning into non-relativistic outflows, potentially unbound winds (even ultra-fast outflows; \citealt{2024MNRAS.527.1424S}), with magnetization fluctuating around $\sigma \sim 1$.
The innermost disconnected magnetic field lines attain regular shape of a double spiral, elevated and converging to a sharp inner tip rooted at the equatorial current layer, reflecting the dominance of $B^r,B^\phi$ components over $B^\theta$ (twisted split monopole).
Such field lines form a rigid structure, rotating at a slow sub-Keplerian rate insensitive to the sense of the BH spin (prograde vs. retrograde).

Plasma motions are well tied to the disconnected magnetic field lines, which channel the accretion flow towards the equatorial current layer.
The inner tips of the disconnected field lines do not choke the accretion flow, which proceeds further due to turbulent cross-field diffusion, channeled along the equatorial current layer towards the horizon, effectively plunging with sub-Keplerian azimuthal velocity $v^\phi$.

In a magnetically saturated BH state, the magnetically disconnected domain acts like a `magnetic insulator' separating two connected magnetospheres `charged' with opposite field lines.
Introduction of plasma-depleted field lines into the disconnected domain leads to its localized geometric thinning.
A critically thin disconnected domain punctures locally, initiating magnetic reconnection between the connected domains and triggering a magnetic flux eruption.

We identified the trigger event, i.e., the first consistent occurrence of relativistic temperature, for a moderate magnetic flux eruption in our reference prograde case. The trigger is a magnetic X-point forming between a band of disconnected lines and a bundle of doubly connected loops.
The frequent and abundant appearance of doubly connected magnetic field loops, short lived features dynamically folding into the BH, is evidence for chaotic disconnection of magnetospheric lines.
In accordance with the principle of self-organized criticality,
only some of those loops trigger eruptions of various magnitudes.

The retrograde case presents additional complication due to a strong shear between the accretion flow and the magnetosphere.
The azimuthal twist of the disconnected magnetic field lines advected by the accretion flow reverses due to spacetime rotation only once they unload their plasma into the BH and become highly magnetized connected field lines.

Magnetic flux eruptions rotate systematically around the BH, this applies not only to the ejected flux tubes (orbiting hotspots at $r \sim 10 M$), but also to their footpoints ($r \sim 2 M$).
In the prograde case, relativistic temperature coincides with the density gap.
In the retrograde case, relativistic temperature appears to corotate with the BH, while density gap appears to corotate with the accretion flow.
The disconnected magnetic field lines rotate much slower than the accretion plasma, both in the prograde and retrograde cases.
The rotation rate of those lines can explain the sub-Keplerian and spin-degenerate rotation speeds of azimuthal patterns imprinted on the simulated BH crescent images, in particular of the flux eruption footpoints.

\begin{acknowledgement}
The authors benefited greatly from discussions with
Marek Sikora, Marek Abramowicz, Mitch Begelman, Beno{\^i}t Cerutti, Jonathan Ferreira, Nicolas Scepi, Alexander Philippov, Jordy Davelaar, Bart Ripperda, Vladimir Zhdankin, Gibwa Musoke.
We gratefully acknowledge Poland's high-performance Infrastructure PLGrid (HPC Centers: ACK Cyfronet AGH, PCSS, CI TASK, WCSS) for providing computer facilities and support within computational grants no. PLG/2023/016444 and PLG/2024/017013;
and the Nicolaus Copernicus Astronomical Center for providing the Chuck cluster.
This work was supported by the Polish National Science Centre grants 2021/41/B/ST9/04306 and 2019/35/B/ST9/04000.
KN acknowledges the hospitality of the Simons Foundation.
\end{acknowledgement}


\begin{appendix}

\section{Magnetic pitch angles of accreting test particles}
\label{app_test_part}

\begin{figure}
\includegraphics[width=\columnwidth,bb=0 0 392 712]{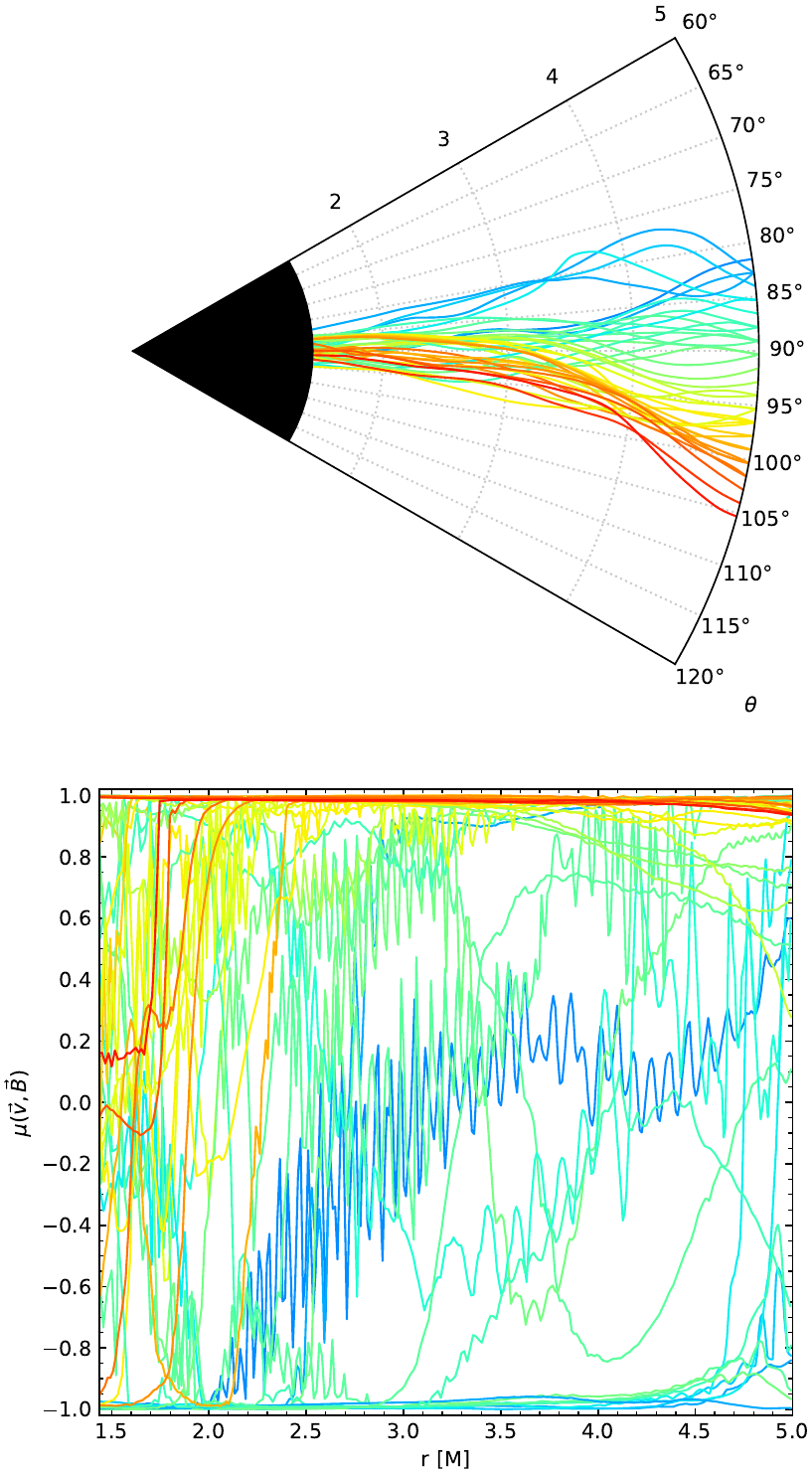}
\caption{Trajectories of test particles in the $(r,\theta)$ projection (upper panel), and cosines of their magnetic pitch angles $\mu(\vec{v},\vec{B})$ as function of $r$ (lower panel).
Line colors indicate the initial latitude $\theta_1$.}
\label{fig_test_particles}
\end{figure}

In order to illustrate channeling of accreting plasma along local magnetic field lines, and diffusion across the field lines, in Figure \ref{fig_test_particles} we show a small representative sample of test plasma particles ($\sim 40$ out of a full sample of $\sim 15000$, in order not to clutter the plot), the trajectories of which were integrated  using the Euler method from 3-linearly interpolated full-resolution velocity field updated every ${\rm d}t = 0.1 M$, starting from their initial positions at $r_1 = 5 M$ for a range of $70^\circ < \theta_1 < 110^\circ$ and $0 < \phi_1 < 360^\circ$ values at $t_1 = 19500 M$ (well before the onset of magnetic flux eruption) to their crossings of the BH horizon at $r_2 = r_{\rm H}$ at $t_2 < 19530 M$.
While all trajectories are smooth in the $(r,\theta)$ plane (and also in the $(r,\phi)$ plane), the cosine of magnetic pitch angle $\mu(\vec{v},\vec{B})$ shows a variety of behaviors.
Most plasma elements (especially those starting away from the accretion midplane) are found to follow their local magnetic field line at small pitch angle ($|\mu| \simeq 1$).
However, some plasma elements show intermediate magnetic pitch angles -- in some cases (those starting close to the midplane) over most of their way, in other cases they switch from following their local field line to diffusion shortly before reaching the horizon.

\end{appendix}


\begin{thebibliography}{}

\bibitem[Ajello et~al.(2015)]{2015ApJ...800L..27A}
Ajello, M., Gasparrini, D., S{\'a}nchez-Conde, M., et~al.\ 2015, \apjl, 800, L27

\bibitem[Aschwanden \& G{\"u}del(2021)]{2021ApJ...910...41A}
Aschwanden, M.~J. \& G{\"u}del, M.\ 2021, \apj, 910, 41

\bibitem[Bak et al.(1988)]{1988PhRvA..38..364B}
Bak, P., Tang, C., \& Wiesenfeld, K.\ 1988, PhRvA, 38, 364

\bibitem[Balbus \& Hawley(1991)]{1991ApJ...376..214B}
Balbus, S.~A. \& Hawley, J.~F.\ 1991, \apj, 376, 214

\bibitem[Begelman et~al.(1984)]{1984RvMP...56..255B}
Begelman, M.~C., Blandford, R.~D., \& Rees, M.~J.\ 1984, RvMP, 56, 255

\bibitem[Begelman et al.(2022)]{2022MNRAS.511.2040B}
Begelman, M.~C., Scepi, N., \& Dexter, J.\ 2022, \mnras, 511, 2040

\bibitem[Biteau et~al.(2020)]{2020NatAs...4..124B}
Biteau, J., Prandini, E., Costamante, L., et~al.\ 2020, NatAs, 4, 124

\bibitem[Blandford \& Znajek(1977)]{1977MNRAS.179..433B}
Blandford, R.~D. \& Znajek, R.~L.\ 1977, \mnras, 179, 433

\bibitem[Blandford et~al.(2019)]{2019ARA&A..57..467B}
Blandford, R., Meier, D., \& Readhead, A.\ 2019, \araa, 57, 467

\bibitem[Camilloni et~al.(2022)]{2022JCAP...07..032C}
Camilloni, F., Dias, O.~J.~C., Grignani, G., et~al.\ 2022, \jcap, 2022, 032

\bibitem[Chakrabarti(1985)]{1985ApJ...288....1C}
Chakrabarti, S.~K.\ 1985, \apj, 288, 1

\bibitem[Chashkina et~al.(2021)]{2021MNRAS.508.1241C}
Chashkina, A., Bromberg, O., \& Levinson, A.\ 2021, \mnras, 508, 1241

\bibitem[Chatterjee \& Narayan(2022)]{2022ApJ...941...30C}
Chatterjee, K. \& Narayan, R.\ 2022, \apj, 941, 30

\bibitem[Conroy et al.(2023)]{2023ApJ...951...46C}
Conroy, N.~S., Baub{\"o}ck, M., Dhruv, V., et~al.\ 2023, \apj, 951, 46

\bibitem[De Villiers \& Hawley(2003)]{2003ApJ...589..458D}
De Villiers, J.-P. \& Hawley, J.~F.\ 2003, \apj, 589, 458

\bibitem[Dexter et~al.(2020)]{2020MNRAS.497.4999D}
Dexter, J., Tchekhovskoy, A., Jim{\'e}nez-Rosales, A., et al.\ 2020, \mnras, 497, 4999

\bibitem[EHT Collaboration(2019.I)]{2019ApJ...875L...1E}
Event Horizon Telescope Collaboration\ 2019, \apjl, 875, L1

\bibitem[EHT Collaboration(2019.IV)]{2019ApJ...875L...4E}
Event Horizon Telescope Collaboration\ 2019, \apjl, 875, L4

\bibitem[EHT Collaboration(2019.V)]{2019ApJ...875L...5E}
Event Horizon Telescope Collaboration\ 2019, \apjl, 875, L5

\bibitem[EHT Collaboration(2019.VI)]{2019ApJ...875L...6E}
Event Horizon Telescope Collaboration\ 2019, \apjl, 875, L6

\bibitem[EHT Collaboration(2021.VII)]{2021ApJ...910L..12E}
Event Horizon Telescope Collaboration\ 2021, \apjl, 910, L12

\bibitem[EHT Collaboration(2021.VIII)]{2021ApJ...910L..13E}
Event Horizon Telescope Collaboration\ 2021, \apjl, 910, L13

\bibitem[EHT Collaboration(2022)]{2022ApJ...930L..12E}
Event Horizon Telescope Collaboration\ 2022, \apjl, 930, L12

\bibitem[EHT Collaboration(2023)]{2023ApJ...957L..20E}
Event Horizon Telescope Collaboration\ 2023, \apjl, 957, L20

\bibitem[Feinstein et al.(2022)]{2022ApJ...925L...9F}
Feinstein, A.~D., Seligman, D.~Z., G{\"u}nther, M.~N., et al.\ 2022, \apjl, 925, L9

\bibitem[Ferreira \& Pelletier(1993)]{1993A&A...276..637F}
Ferreira, J. \& Pelletier, G.\ 1993, \aap, 276, 637

\bibitem[Fishbone \& Moncrief(1976)]{1976ApJ...207..962F}
Fishbone, L.~G. \& Moncrief, V.\ 1976, \apj, 207, 962

\bibitem[Gammie et al.(2003)]{2003ApJ...589..444G}
Gammie, C.~F., McKinney, J.~C., \& T{\'o}th, G.\ 2003, \apj, 589, 444

\bibitem[Genestreti et~al.(2023)]{2023JGRA..12831758G}
Genestreti, K.~J., Farrugia, C.~J., Lu, S., et~al.\ 2023, JGRA, 128, e2023JA031758

\bibitem[Gezari(2021)]{2021ARA&A..59...21G}
Gezari, S.\ 2021, \araa, 59, 21

\bibitem[Ghisellini et~al.(2014)]{2014Natur.515..376G}
Ghisellini, G., Tavecchio, F., Maraschi, L., et~al.\ 2014, \nat, 515, 376

\bibitem[Goddi et al.(2021)]{2021ApJ...910L..14G}
Goddi, C., Mart{\'\i}-Vidal, I., Messias, H., et al.\ 2021, \apjl, 910, L14

\bibitem[GRAVITY Collaboration(2018)]{2018A&A...618L..10G}
GRAVITY Collaboration\ 2018, \aap, 618, L10

\bibitem[Hada et~al.(2016)]{2016ApJ...817..131H}
Hada, K., Kino, M., Doi, A., et~al.\ 2016, \apj, 817, 131

\bibitem[Hakobyan et al.(2023)]{2023ApJ...943L..29H}
Hakobyan, H., Ripperda, B., \& Philippov, A.~A.\ 2023, \apjl, 943, L29

\bibitem[Igumenshchev(2008)]{2008ApJ...677..317I}
Igumenshchev, I.~V.\ 2008, \apj, 677, 317

\bibitem[James et~al.(2024)]{2024A&A...687A.185J}
James, B., Janiuk, A., \& Karas, V.\ 2024, \aap, 687, A185

\bibitem[Janiuk \& James(2022)]{2022A&A...668A..66J}
Janiuk, A. \& James, B.\ 2022, \aap, 668, A66

\bibitem[Jiang et~al.(2021)]{2021NatAs...5.1126J}
Jiang, C., Feng, X., Liu, R., et~al.\ 2021, NatAs, 5, 1126

\bibitem[Junor et~al.(1999)]{1999Natur.401..891J}
Junor, W., Biretta, J.~A., \& Livio, M.\ 1999, \nat, 401, 891

\bibitem[Kellermann et~al.(1998)]{1998AJ....115.1295K}
Kellermann, K.~I., Vermeulen, R.~C., Zensus, J.~A., et~al.\ 1998, \aj, 115, 1295

\bibitem[Kim et~al.(2018)]{2018A&A...616A.188K}
Kim, J.-Y., Krichbaum, T.~P., Lu, R.-S., et~al.\ 2018, \aap, 616, A188

\bibitem[Komissarov(2004)]{2004MNRAS.350..427K}
Komissarov, S.~S.\ 2004, \mnras, 350, 427

\bibitem[Komissarov \& McKinney(2007)]{2007MNRAS.377L..49K}
Komissarov, S.~S. \& McKinney, J.~C.\ 2007, \mnras, 377, L49

\bibitem[Kulkarni \& Romanova(2008)]{2008MNRAS.386..673K}
Kulkarni, A.~K. \& Romanova, M.~M.\ 2008, \mnras, 386, 673

\bibitem[Lasota et~al.(2014)]{Lasota2014}
Lasota, J.-P., Gourgoulhon, E., Abramowicz, M., et~al.\ 2014, PhRvD, 89, 024041

\bibitem[Lazarian \& Vishniac(1999)]{1999ApJ...517..700L}
Lazarian, A. \& Vishniac, E.~T.\ 1999, \apj, 517, 700

\bibitem[Liska et~al.(2020)]{2020MNRAS.494.3656L}
Liska, M., Tchekhovskoy, A., \& Quataert, E.\ 2020, \mnras, 494, 3656

\bibitem[Liska et~al.(2022)]{Liska22}
Liska, M.~T.~P., Musoke, G., Tchekhovskoy, A., et~al.\ 2022, \apjl, 935, L1

\bibitem[LHAASO Collaboration(2023)]{2023Sci...380.1390L}
LHAASO Collaboration\ 2023, Science, 380, 1390

\bibitem[Lu \& Hamilton(1991)]{1991ApJ...380L..89L}
Lu, E.~T. \& Hamilton, R.~J.\ 1991, \apjl, 380, L89

\bibitem[Lu et al.(2023)]{2023Natur.616..686L}
Lu, R.-S., Asada, K., Krichbaum, T.~P., et al.\ 2023, \nat, 616, 686

\bibitem[Lubow et al.(1994)]{1994MNRAS.267..235L}
Lubow, S.~H., Papaloizou, J.~C.~B., \& Pringle, J.~E.\ 1994, \mnras, 267, 235

\bibitem[Lynden-Bell(1969)]{1969Natur.223..690L}
Lynden-Bell, D.\ 1969, \nat, 223, 690

\bibitem[Madejski \& Sikora(2016)]{2016ARA&A..54..725M}
Madejski, G. \& Sikora, M.\ 2016, \araa, 54, 725

\bibitem[Manikantan et~al.(2024)]{2024ApJ...965..175M}
Manikantan, V., Kaaz, N., Jacquemin-Ide, J., et~al.\ 2024, \apj, 965, 175

\bibitem[Marshall et~al.(2018)]{2018MNRAS.478.1837M}
Marshall, M.~D., Avara, M.~J., \& McKinney, J.~C.\ 2018, \mnras, 478, 1837

\bibitem[McKinney et~al.(2012)]{2012MNRAS.423.3083M}
McKinney, J.~C., Tchekhovskoy, A., \& Blandford, R.~D.\ 2012, \mnras, 423, 3083

\bibitem[Mirabel \& Rodr{\'\i}guez(1999)]{1999ARA&A..37..409M}
Mirabel, I.~F. \& Rodr{\'\i}guez, L.~F.\ 1999, \araa, 37, 409

\bibitem[Najafi-Ziyazi et al.(2024)]{2024MNRAS.531.3961N}
Najafi-Ziyazi, M., Davelaar, J., Mizuno, Y., et al.\ 2024, \mnras, 531, 3961

\bibitem[Nalewajko et~al.(2020)]{2020A&A...634A..38N}
Nalewajko, K., Sikora, M., \& R{\'o}{\.z}a{\'n}ska, A.\ 2020, \aap, 634, A38

\bibitem[Nalewajko(2023)]{2023mgm..conf..339N}
Nalewajko, K.\ 2023, The Sixteenth Marcel Grossmann Meeting, 339

\bibitem[Nalewajko et~al.(2024)]{2024hepr.conf}
Nalewajko, K., Kapusta, M., \& Janiuk, A.\ 2024, High Energy Phenomena in Relativistic Outflows VIII

\bibitem[Narayan et al.(2003)]{2003PASJ...55L..69N}
Narayan, R., Igumenshchev, I.~V., \& Abramowicz, M.~A.\ 2003, \pasj, 55, L69

\bibitem[Narayan et~al.(2022)]{2022MNRAS.511.3795N}
Narayan, R., Chael, A., Chatterjee, K., et~al.\ 2022, \mnras, 511, 3795

\bibitem[Noble et~al.(2006)]{2006ApJ...641..626N}
Noble, S.~C., Gammie, C.~F., McKinney, J.~C., et~al.\ 2006, \apj, 641, 626

\bibitem[Parfrey \& Tchekhovskoy(2023)]{2023arXiv231104291P}
Parfrey, K. \& Tchekhovskoy, A.\ 2023, {\apj} submitted, arXiv:2311.04291

\bibitem[Piran(2004)]{2004RvMP...76.1143P}
Piran, T.\ 2004, RvMP, 76, 1143

\bibitem[Porth et~al.(2019)]{2019ApJS..243...26P}
Porth, O., Chatterjee, K., Narayan, R., et~al.\ 2019, \apjs, 243, 26

\bibitem[Porth et~al.(2021)]{2021MNRAS.502.2023P}
Porth, O., Mizuno, Y., Younsi, Z., et~al.\ 2021, \mnras, 502, 2023

\bibitem[Punsly et~al.(2009)]{2009ApJ...704.1065P}
Punsly, B., Igumenshchev, I.~V., \& Hirose, S.\ 2009, \apj, 704, 1065

\bibitem[Punsly \& Chen(2021)]{2021ApJ...921L..38P}
Punsly, B. \& Chen, S.\ 2021, \apjl, 921, L38

\bibitem[Ripperda et~al.(2022)]{2022ApJ...924L..32R}
Ripperda, B., Liska, M., Chatterjee, K., et~al.\ 2022, \apjl, 924, L32

\bibitem[Scepi et~al.(2022)]{2022MNRAS.511.3536S}
Scepi, N., Dexter, J., \& Begelman, M.~C.\ 2022, \mnras, 511, 3536

\bibitem[Scepi et~al.(2024)]{2024MNRAS.527.1424S}
Scepi, N., Begelman, M.~C., \& Dexter, J.\ 2024, \mnras, 527, 1424

\bibitem[Shakura \& Sunyaev(1973)]{1973A&A....24..337S}
Shakura, N.~I. \& Sunyaev, R.~A.\ 1973, \aap, 24, 337

\bibitem[Stone \& Gardiner(2007)]{2007ApJ...671.1726S}
Stone, J.~M. \& Gardiner, T.\ 2007, \apj, 671, 1726

\bibitem[Stone et~al.(2020)]{2020ApJS..249....4S}
Stone, J.~M., Tomida, K., White, C.~J., et~al.\ 2020, \apjs, 249, 4

\bibitem[Tchekhovskoy et~al.(2010)]{Tchekhovskoy2010}
Tchekhovskoy A., Narayan R., McKinney J.~C.\ 2010, \apj, 711, 50

\bibitem[Tchekhovskoy et~al.(2011)]{2011MNRAS.418L..79T}
Tchekhovskoy, A., Narayan, R., \& McKinney, J.~C.\ 2011, \mnras, 418, L79

\bibitem[Tchekhovskoy \& McKinney(2012)]{2012MNRAS.423L..55T}
Tchekhovskoy, A. \& McKinney, J.~C.\ 2012, \mnras, 423, L55

\bibitem[Urry \& Padovani(1995)]{1995PASP..107..803U}
Urry, C.~M. \& Padovani, P.\ 1995, \pasp, 107, 803

\bibitem[Walker et~al.(2018)]{2018ApJ...855..128W}
Walker, R.~C., Hardee, P.~E., Davies, F.~B., et~al.\ 2018, \apj, 855, 128

\bibitem[Werner \& Uzdensky(2017)]{2017ApJ...843L..27W}
Werner, G.~R. \& Uzdensky, D.~A.\ 2017, \apjl, 843, L27

\bibitem[White et~al.(2016)]{2016ApJS..225...22W}
White, C.~J., Stone, J.~M., \& Gammie, C.~F.\ 2016, \apjs, 225, 22

\bibitem[White et~al.(2019)]{2019ApJ...874..168W}
White, C.~J., Stone, J.~M., \& Quataert, E.\ 2019, \apj, 874, 168

\bibitem[Yuan \& Narayan(2014)]{2014ARA&A..52..529Y}
Yuan, F. \& Narayan, R.\ 2014, \araa, 52, 529

\bibitem[Zhang(2018)]{2018pgrb.book.....Z}
Zhang, B.\ 2018, The Physics of Gamma-Ray Bursts, Cambridge Univeristy Press

\bibitem[Zhang et~al.(2024)]{2024ApJ...962..135Z}
Zhang, G.-Q., B{\'e}gu{\'e}, D., Pe'er, A., et al.\ 2024, \apj, 962, 135

\bibitem[Zhdankin et~al.(2023)]{2023PhRvR...5d3023Z}
Zhdankin, V., Ripperda, B., \& Philippov, A.~A.\ 2023, PhRvR, 5, 043023

\bibitem[Zhu et~al.(2024)]{2024MNRAS.528.2883Z}
Zhu, Z., Stone, J.~M., \& Calvet, N.\ 2024, \mnras, 528, 2883

\end{thebibliography}
\end{document}